\documentclass[12pt,preprint]{aastex}
\usepackage{graphicx}
\usepackage{natbib}
\newcommand{\michelmu}{\mu}
\newcommand{\vchar}{\beta_{\rm c}}
\newcommand{\vfast}{\beta_{\rm f}}
\newcommand{\phat}{\hat{p}}
\newcommand{\lhat}{\hat{L}}

\newcommand{\Owind}{\Omega_{\rm w}}
\newcommand{\gheat}{\hat{\gamma}}
\newcommand{\binfty}{B'_0}
\newcommand{\ncold}{n_{\rm c}}
\newcommand{\eps}{\varepsilon}
\newcommand{\unitphi}{\vec{\hat{\phi}}}
\newcommand{\unitr}{\vec{\hat{r}}}

\newcommand{\unittheta}{\vec{\hat{\theta}}}

\newcommand{\eqb}{\begin{eqnarray}}
\newcommand{\eqe}{\end{eqnarray}}
\newcommand{\diff}{{\rm d}}

\newcommand{\rlight}{r_{\rm L}}

\newcommand{\gtear}{\gamma_{\rm t}}

\newcommand{\ldebye}{\lambda_{\rm D}}

\newcommand{\sech}{{\rm sech}}
\newcommand{\trelax}{t_{\rm coll}}
\newcommand{\tdyn}{t_{\rm dyn}}
\newcommand{\tbohm}{t_{\rm B}}

\newcommand{\nc}{\newcommand}
\nc{\lab}{\label}
\nc{\bea}{\begin{eqnarray}}
\nc{\eea}{\end{eqnarray}}
\nc{\be}{\begin{equation}}
\nc{\ee}{\end{equation}}

\nc{\period}{P}
\nc{\pdot}{\dot P}
\nc{\rdiss}{r_d}
\nc{\rshock}{r_s}
\nc{\params}{$\{\Delta,\xi,\gamma/\mu \}$}
\nc{\permutation}{$\{\alpha_1,\alpha_2,\alpha_3\}$}
\nc{\equalparams}{$\alpha_1\simeq \alpha_2\simeq \alpha_3$}
\nc{\Ndot}{\dot N}
\nc{\gam}{\gamma}
\nc{\rcormich}{r}
\nc{\rtear}{r}
\nc{\rfast}{r}
\nc{\rloss}{r}
\nc{\omegawind}{\Omega_{\rm w}}
\nc{\f}{\frac}
\nc{\pressure}{p}
\nc{\rchar}{r_d}
\nc{\lft}{\left}
\nc{\rgt}{\right}
\nc{\non}{\nonumber}
\nc{\sh}{\sinh}
\nc{\ch}{\cosh}
\nc{\muode}{\eta}
\nc{\Bp}{B'}
\nc{\rratio}{\rdiss/\rshock}
\nc{\rslow}{R}

\shorttitle{Dissipation in Poynting-flux dominated flows}
\shortauthors{Kirk \& Skj\ae raasen}

\begin{document}
\title{Dissipation in Poynting-flux Dominated Flows:\\
 the $\sigma$-Problem of the Crab Pulsar Wind}
\author{J. G. Kirk and O. Skj\ae raasen}
\affil{Max-Planck-Institut f\"ur Kernphysik, Postfach 103980, D-69029 Heidelberg, Germany}
\email{john.kirk@mpi-hd.mpg.de\ olaf.skjaeraasen@mpi-hd.mpg.de}
\date{Received \dots}
\begin{abstract}
Flows in which energy is transported predominantly as Poynting flux 
are thought to occur in pulsars, gamma-ray bursts and 
relativistic jets from compact objects. 
The fluctuating component of the magnetic field in such a flow
can in principle be dissipated by magnetic 
reconnection, and used to accelerate the flow. We investigate how rapidly
this transition can take place, by implementing 
into a global MHD 
model,
that uses a thermodynamic description of the plasma,
explicit, physically motivated 
prescriptions for the dissipation rate:
a lower limit on this rate is given by limiting the maximum drift speed of the current
carriers to that of light, an upper limit follows from demanding that the
dissipation zone expand only subsonically in the comoving frame and a further prescription is
obtained by assuming that the expansion speed is limited by the growth rate of the 
relativistic tearing mode. 
In each case, solutions are presented which give the Lorentz factor of a spherical
wind containing a transverse, oscillating magnetic field component 
as a function of radius. In the
case of the Crab pulsar, we find that the Poynting flux can be dissipated
before the wind reaches the inner edge of the Nebula if 
the pulsar emits electron positron pairs
at a rate $\dot{N}_\pm>10^{40}\,{\rm s}^{-1}$, 
thus providing a possible solution to the \lq\lq $\sigma$-problem\rq\rq.
\end{abstract}

         
\maketitle

\section{Introduction}
Powerful outflows in which energy is transported predominantly as Poynting
flux rather than kinetic energy arise in several astrophysical contexts. These
include pulsars \citep{gunnostriker69,reesgunn74}, gamma-ray bursts 
\citep{usov92,spruit99,lyutikovblackman01}, 
and the jets from galactic compact objects
and active galactic nuclei \citep{lovelaceetal02}.
An important generic property of such flows is that they are highly
relativistic. 
Force-free solutions, valid in the inner regions, indicate that radial
flows accelerate quickly \citep{buckley77,contopouloskazanas02} until
they cross the fast magnetosonic point, at which their
bulk Lorentz factor $\gamma$ satisfies $\gamma=\sqrt{\sigma}$. 
The parameter $\sigma$, which is
the ratio of Poynting flux 
to particle-born energy flux is by definition large under these conditions. 
Eventually, such flows release their energy into particles and/or radiation.
In principle, this can happen if the flow diverges sufficiently rapidly and
is thus forced to accelerate. However,
this occurs only very slowly for ideal relativistic MHD flows which are
initially radial 
\citep{beskinkuznetsovarafikov98,chiuehlibegelman98,bogovalovtsinganos99,bogovalov01,lyubarskyeichler01,lyubarsky02a}.
In the case of non-steady winds, such as those driven by an obliquely
rotating magnetic dipole, the current required to support the oscillating component
of the magnetic field decreases more slowly with radius than the density of available
charge carriers \citep{usov75,melatosmelrose96}. Consequently, 
the ideal MHD approximation must break down at some point in the
outflow where the implied minimum drift speed reaches that of light.
However, internal dissipative effects 
may intervene before this stage is reached, reducing the wave amplitude
and decreasing the required current. 
Shock fronts, which provide efficient dissipation for low $\sigma$ flows, are 
rather slow to dissipate the energy of a Poynting-flux dominated outflow,
although they may be effective if the particle density is high
\citep{lyubarsky02b}. An alternative is dissipation by annihilation or
reconnection of the oscillating part of the 
magnetic field \citep{coroniti90,michel94}, and this is the process we 
investigate
in detail in this paper. The effectiveness of such a dissipation mechanism 
is an important
ingredient in models of pulsed high energy radiation from pulsar winds
\citep{kirkskjaeraasengallant02,skjaeraasenetalboston02,skjaeraasenkirk01} 
and the after-glows of gamma-ray bursts
\citep{drenkhahnspruit02}. 

Although now invoked in models of AGN jets and gamma-ray bursts
\citep{thompson94,schopperetal98,heinzbegelman00,spruitdaignedrenkhahn01}, the 
original motivation for considering reconnection in Poynting-flux dominated
outflows was the so-called \lq\lq$\sigma$-problem\rq\rq.
This term is generally used to refer to a
puzzle posed by the Crab Nebula and the pulsar wind which powers
it: 
according to estimates of the number of pairs emitted by the pulsar 
\citep{arons83,hibschmanarons01b,gallantetal02}, the energy flux 
carried by the wind whilst close to the star is dominated by Poynting flux
\citep{michel82}. 
However, at the inner edge of the Crab Nebula, where 
the wind passes through a termination shock, it must have a very low
magnetization, $\sigma\sim10^{-3}$, if it is to match the observed expansion
speed of the nebula 
\citep{reesgunn74,kennelcoroniti84,emmeringchevalier87} and the axial ratio of its
optical isophotes \citep{begelmanli92}. Thus, unless the 
structure of the termination shock is very different to that of a hydrodynamic
shock, dissipation of the
magnetic field in the flow must occur before the flow reaches this 
radius. 

\begin{figure}
\epsscale{0.4}
\plotone{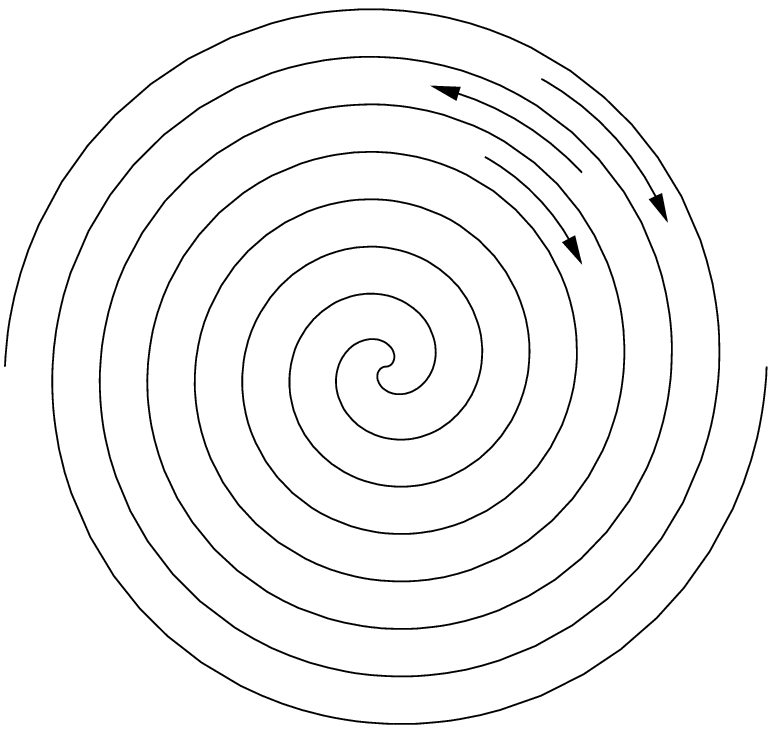}\\
\epsscale{0.5}
\plotone{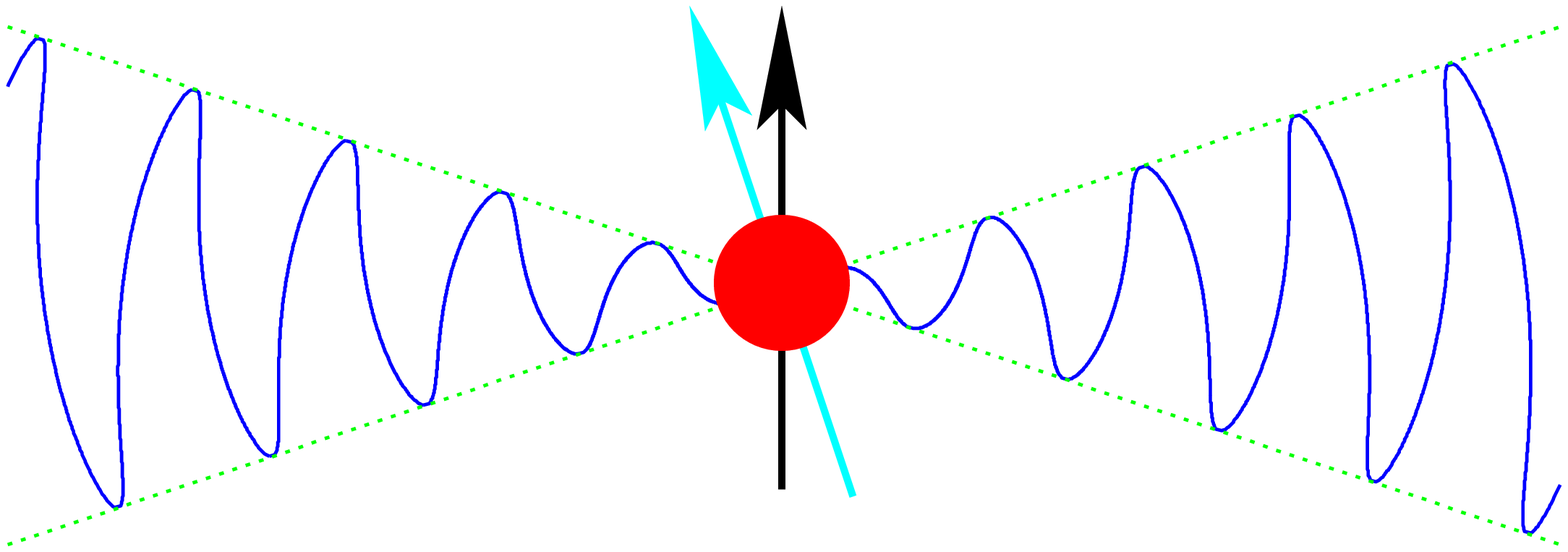}
\caption{Top: The Parker spiral pattern in the equatorial plane. The arrows indicate the magnetic
field direction of an obliquely rotating magnetic dipole, and the solid line gives the 
position of the current sheet. Bottom: the poloidal structure of the current sheet 
(thick, solid line). The arrows denote the spin axis (vertical) and magnetic
dipole axis (inclined), and the thin line bounds the equatorial
zone of the wind, in which magnetic reconnection is possible.}
\label{fig:parkerspiral}
\end{figure}

The wind of the Crab pulsar is driven by a rotating neutron star in which a
non-axisymmetric magnetic field is embedded --- often assumed to be an oblique
magnetic dipole. Close to the equatorial plane, the magnetic field in the wind
forms toroidal \lq\lq stripes\rq\rq\ of opposite polarity, separated by a
current sheet \citep{coroniti90,bogovalov99}, as illustrated in Fig.~\ref{fig:parkerspiral}. 
In principle, a dissipative process
operating on the scale of the wavelength of the striped pattern is capable of 
annihilating the oscillating component, which, in the
equatorial plane, corresponds to the entire magnetic field.
But the rate at which dissipation proceeds is uncertain.
Using a particular prescription in which the speed of the charge carriers
attains the maximum value of $c$, 
\citet{coroniti90} and \citet{michel94}
found for the case of the Crab pulsar that
dissipation should be completed well inside the termination shock, which
occurs at a radius of roughly $10^{9}\rlight$ (where $\rlight=c/\Omega$ is the radius
of the light-cylinder and $\Omega$ the angular velocity of the star). 
However, an important effect of dissipation 
is that it causes the flow to accelerate. 
As magnetic field energy
is converted into heat, the plasma performs work on the expanding wind 
and converts the heat into kinetic energy. 
Because of the relativistic dilation of time, this effect, 
which was not included in the treatments 
of \citet{coroniti90} and \citet{michel94}, leads to an apparent decrease in the 
rate of dissipation in the rest frame of the star
[Lyubarsky \& Kirk (2001)\nocite{lyubarskykirk01} henceforth LK]. 
As a result, this rate of dissipation does not 
solve the $\sigma$-problem.

Physically, this approach to dissipation is based on a one-dimensional picture in that it 
corresponds to magnetic field annihilation in a current sheet that is uniform in
the plane perpendicular to the radius vector. In reality, we might expect the sheet to 
break up into a large number of regions where
dissipation proceeds via magnetic reconnection in flow patterns which are 
two or three dimensional in nature (see, for example Priest and
Forbes~2000\nocite{priestforbes00}). 
In this case, the global effect on the flow can be found by 
(i)
maintaining the
assumption that 
dissipation is confined to a layer around the field reversal, but allowing the
thickness of this layer to be determined by an estimate of the average dissipation
rate based on the physics of reconnection 
and (ii) assigning to the layer an average energy density and pressure,
related to each other by an effective equation of state.
The prescription used by \citet{coroniti90}, \citet{michel94}
and \citet{lyubarskykirk01} is equivalent to setting the thickness of the
dissipation layer equal to the gyro radius of the current carriers
and using the equation of state of an ideal relativistic gas.
This gives
a reasonable lower limit, but other
prescriptions have been proposed \citep{lyubarsky96,drenkhahn02,lyutikovuzdensky02}.
In this paper we argue that an
{\em upper} limit to the reconnection rate can be implemented by 
limiting the expansion speed of the layer to that of sound in an
unmagnetized gas. We present the global properties of reconnecting, 
spherically symmetric, relativistic winds,
including the dependence of the Lorentz factor on radius,
for each prescription. 
Our findings apply to any spherical relativistic flow that contains 
a toroidal field with radial fluctuations, but we phrase 
the discussion in terms of a pulsar wind, since
this fixes the wavelength of the fluctuation as well as its initial 
amplitude. In the particular case of the 
Crab pulsar, we show that the $\sigma$-problem can be resolved 
if the initial value (i.e, the value at the fast magnetosonic point) 
of this parameter is at most 
$10^3$, or, equivalently, if the $\michelmu$-parameter [see Eq.~(\ref{eq:mu_michel})]
introduced by \citet{michel69} is less than a few times $10^4$.

The paper is organized as follows:
in Sect.~\ref{mhdwinds} we review the equations derived by
\citet{lyubarskykirk01} that describe, in the short
wavelength limit, spherically
symmetric MHD winds containing an oscillating component of the magnetic
field which can undergo damping by dissipative effects.
In Sect.~\ref{sheet} we discuss the modelling of the 
dissipative layer and describe how it can be incorporated into
the
MHD description. Section~\ref{reconnection} contains a discussion of the
various prescriptions for dissipation. 
In section \ref{results},
asymptotic solutions, valid for relativistic plasma temperatures,
are found that give good approximations to the   
global properties of the flows, such as the
radius at which dissipation is complete.
These are compared to numerical solutions, which indicate the
sensitivity to variations in the initial conditions, delineate regions
in which the dissipation prescriptions become unphysical
and quantify departures from the asymptotic solutions when the plasma
temperature is nonrelativistic. 
We compute the effective diffusivity implied by our results in
Sect.~\ref{diffusivity} and compare it to the Bohm diffusivity. 
The implications of our results, in particular for the Crab pulsar and its nebula, 
are discussed in Sect.~\ref{discussion} and a summary of our conclusions is
presented in Sect.~\ref{conclusions}.

\section{MHD equations in the short wavelength approximation}
\label{mhdwinds}
In this paper we investigate the evolution of 
a radial MHD wind containing a purely toroidal magnetic
field.
This idealisation is thought to be a reasonably good approximation 
in the case of a pulsar for radii far outside the light
cylinder, $r\gg\rlight$ [see, for example \citet{coroniti90}]. 
However, alternative pictures in which the radial component of the magnetic field plays
a role are possible \citep{kuijpers01}. 
Following \citet{lyubarskykirk01} we
consider the slow evolution of a periodic entropy wave, containing two
reversals of the magnetic direction per oscillation, as it propagates outwards in
the spherically symmetric wind. 
The flow pattern is assumed to be steady in the sense that at fixed radius the
amplitude of the wave is constant in time, 
so that the flow variables are exactly periodic. The radial evolution 
is governed by two factors: the gradual annihilation of 
the oscillating component of the magnetic field embedded in the wave according 
to one of the microphysical prescriptions discussed in Sect.~\ref{reconnection} 
and the expansion and acceleration of the plasma as a result of the spherical 
geometry and the outwardly directed force arising from the 
pressure gradient. The evolution of the wave is assumed to be slow ---
significant changes are permitted only over radial scales large compared to
the wavelength ($=2\pi\rlight v/c$, where $v$ is the bulk radial velocity).
The equations describing the evolution are 
obtained from a perturbation expansion in the small 
parameter $\rlight/r$ and 
involves averaging over a wave period the equations of
continuity, energy and entropy.
The procedure is described in detail in LK. 
The zeroth order equations give the familiar conditions for the 
entropy wave mode: 
pressure equilibrium between plasma and magnetic field and
zero bulk speed in the wave frame. Imposing non-secularity conditions 
on the first order equations then gives expressions governing the slow 
evolution of the zeroth order quantities.
From the continuity equation one obtains:
\eqb
{\partial\over\partial r}\left(
r^2\gamma v\left<n'\right>
\right)
&=&0,
\label{continuity}
\eqe
and from the energy equation:
\eqb
{\partial\over\partial r}\left(
r^2\gamma^2 v\left<e'+p'+{B^2\over 4\pi\gamma^2}\right>
\right)
&=&0.
\label{energy}
\eqe
The entropy equation 
includes ohmic dissipation to first order in $\rlight/r$, and can be written  
in terms of the magnetic field, using Maxwell's equations:
\eqb
{\left< e'+p'\right>\over r^2}{\partial \over \partial 
r}\left(r^2\gamma v\right)
&&
\nonumber\\
+\gamma v{\partial\over\partial r}\left<e'\right>
+{1\over4\pi\gamma r}
\left<B{\partial\over\partial r}\left(r vB\right)\right>
&=&0.
\label{entropy}
\eqe
In these expressions, gaussian units have been employed, the magnetic 
field and fluid velocity 
in the laboratory frame (in spherical polar coordinates) are 
$\vec{B}=B(r,t)\unitphi$ and $\vec{v}=v(r,t)\unitr$
and the notation 
$\left<\cdot\right>$ denotes an average over a wave period.
The proper particle density is denoted by $n'$, the proper energy density $e'$
is the component $T^{0,0}$ of the stress-energy tensor 
measured in the rest frame 
and the pressure $p'$ 
refers to the radial component $T^{1,1}$ in the same frame.

\section{The current sheet}
\label{sheet}
In order to perform the averaging in Eqs.~(\ref{continuity}), (\ref{energy})
and (\ref{entropy}) an explicit model of the entropy wave and the associated
current is needed. \citet{lyubarskykirk01} used a fluid model in which a hot,
unmagnetized plasma separates regions of opposing magnetic field containing 
cold plasma. The transition at the interface
of the hot and cold regions was assumed to be sharp. 
In this paper, we use instead an exact stationary solution of the full
Vlasov/Maxwell kinetic equations that contains a magnetic field reversal. 
This solution also forms the basis of the discussion of the microphysics 
presented in Sect.~\ref{reconnection}. When an average over one wavelength is 
performed, the equations obtained are essentially the same as those used by 
\citet{lyubarskykirk01}, with two minor differences. The first is connected
with the presence of a cold plasma component within the hot parts of the
sheet and is noted below where it appears. The second arises when 
the drift speed of the charge carriers is
relativistic and it becomes necessary to compute the effective ratio of
specific heats of the plasma as a function of position according to 
Eq.~(\ref{effspecificheat}). We expect this effect to be unimportant and 
do not investigate it in detail.
The use of an equilibrium distribution presupposes an effective
collision time $\trelax$ that is short compared to the
dynamical time $\tdyn$. On the other hand, each dissipation
prescription has an associated value of $\trelax$. In
Sect.~\ref{diffusivity} we show that indeed $\trelax<\tdyn$ in each
case.

The exact stationary solution we use was originally found by \citet{harris62}
for the nonrelativistic case
and subsequently generalized to apply also to the case of 
relativistic particle temperatures and drift speeds by \citet{hoh68}.
It contains three parameters which can be chosen to
be the magnetic field $\binfty$ far from the current sheet, the number density
of particles of each charge $N'_\pm$ at the center of the sheet and the sheet
width $a$, each measured in a frame of reference in which the sheet is
stationary and the electric field vanishes.
One can then 
define a temperature $T$ (in energy units) which is independent of 
position:
\eqb
T&=&{{\binfty}^2\over 16\pi N'_\pm}.
\eqe
The Debye length at the center of the sheet
\eqb
\ldebye&=&\left(T\over 4\pi N'_\pm e^2\right)^{1/2},
\eqe
is, to within a factor of the order of unity, equal to the gyro radius of
a particle of energy $T$ in the magnetic field $\binfty$,
provided $T\gg mc^2$. The sheet is neutral and 
the electric field vanishes everywhere. The current is carried by oppositely 
directed drifts of oppositely charged particles. Specialising to the case in 
which only positrons and electrons are present, the speed $\pm c\beta$ with 
which each component drifts is
\eqb
\beta&=&{\ldebye\over a},
\label{driftspeed}
\eqe
and is independent 
of position in the sheet.
The current sheet in this solution is planar, but can be used to 
describe each of the sequence of nested
quasi-spherical sheets contained in the radial wind, provided the sheet thickness
remains small compared to the wavelength. The frame of reference used
to portray the
sheet solution is one which moves radially outwards with the
bulk velocity of the wind. A derivation of the internal structure is presented
in Appendix~\ref{harris}. For a single sheet located at radius $r$ at
$t=0$, the
magnetic field $B'(r)$ and the density of each species of charge carrier $n'_\pm$ are
\eqb
B'(r,t)&=&\binfty(r)\tanh\left[{\gamma v t\over a(r)}\right],
\label{bsheet}\\
n'_\pm(r,t)&=&N'_\pm(r)\sech^2\left[{\gamma v t\over a(r)}\right],
\label{nsheet}
\eqe
An additional spatially uniform component of cold, 
neutral, current-free plasma of arbitrary density can be added to this solution 
without changing the above results. The solution is also unaffected by 
the addition of a constant magnetic field of arbitrary strength in the 
$\unittheta$ direction. 

One wavelength of the entropy wave contains two current sheets of opposite 
polarity. Defining the fraction $\Delta$ of the wave occupied by hot plasma 
as
\eqb
\Delta&=&{2ac\over\gamma\pi\rlight v},
\eqe
and averaging over one wave period, one finds for the particle number density
$n'$ (=$2n'_\pm$):
\eqb
\left<n'\right>&=&2N'_\pm\Delta+\ncold',
\eqe
where $\ncold'$ is the density of the uniform cold background plasma. 
The average of the pressure $p'$ follows from the stress-energy tensor of the
sheet particles (Eq.~\ref{stress11})
\eqb
\left<p'\right>&=&2N'_\pm\Delta T\,=\,{{\binfty}^2\over 8\pi}\Delta,
\eqe
and this can be related to the average energy density by
defining the effective ratio of specific heats $\gheat$
according to 
\eqb
p'&=&\left(\gheat-1\right)\left(e'-n'mc^2\right),
\label{specheats}
\eqe  
[see Eq.~(\ref{effspecificheat})]. 
In principle, $\gheat$ is capable of accounting not only for potential
anisotropies in the gas ($p'$ refers to the $r$--$r$ component of the
stress-energy tensor --- see Appendix~\ref{harris}) but also for other
non-equilibrium particle distributions, e.g., those of power-law type.

The magnetic field as given by Eq.~(\ref{bsheet}) leads to 
\eqb
\left<{B^2\over4\pi\gamma^2}\right>&=&
\left<{{B'}^2\over 4\pi}\right>,
\nonumber\\
&=&{{\binfty}^2\over4\pi}\left(1-\Delta\right),
\\
\left<B{\partial\over\partial r}\left(r v B\right)\right>&=&
\gamma^2{\binfty}^2\left(1-\Delta\right){\partial\over\partial r}
\left(r v\right)+
\nonumber\\
&&{r v\over 2}{\partial\over\partial r}\left[
{\gamma^2\binfty}^2\left(1-\Delta\right)\right].
\eqe

Using these expressions, the (integrated) continuity and energy 
equations and the entropy equation (\ref{continuity}, \ref{energy}
and \ref{entropy}) become:
\eqb
r^2\gamma v\left(\ncold' +
2N'_\pm \Delta\right)\,=\,{2\dot{N}_\pm\over\Owind}&\equiv&\Phi,
\label{continuity2} \lab{eq:cont2}\\
\gamma m c^2\Phi + r^2\gamma^2 v{{\binfty}^2\over 8\pi}
\left[2+\left({2-\gheat\over\gheat-1}
\right)\Delta\right]\,=\,
{L\over\Owind}&\equiv&\Psi,
\label{energy2} \lab{eq:energy2}\\
\left({\gheat\over\gheat-1}\right)
{{\binfty}^2\Delta\over 8\pi r^2}{\partial\over\partial r}
\left(r^2 \gamma v\right)
+
&&\nonumber\\
\gamma v{\partial\over\partial r}\left({\Delta\over\gheat-1} {{\binfty}^2
\over8\pi}\right)
+
{v\over\gamma}{\partial\over\partial r}
\left[\gamma^2\left(1-\Delta\right){{\binfty}^2\over8\pi}\right]
&&\nonumber\\
+{\gamma {\binfty}^2 \left(1-\Delta\right)\over 4\pi 
r}{\partial\over\partial r}
\left(r v\right)
&=&0,
\label{entropy2}    \lab{eq:entropy2}
\eqe
where $\dot{N}_\pm$ is the rate at which pairs leave the star 
(corresponding to $2\dot{N}_\pm$ particles per second)
and $L$ is the luminosity in the wind,
which is assumed to occupy a solid angle $\Owind$.
The constants 
$\Phi$ and $\Psi$ represent the particle flux and the total energy flux.

These equations are complemented by the condition that far from the center of
the current sheets the magnetic flux remain frozen into the plasma:
\eqb
{\binfty\over r\ncold'}&=&{\rm constant}.
\label{flux2} \lab{eq:fluxfreeze2}
\eqe

We now define
the fraction $\xi$ of particles carried in the hot plasma:
\eqb
\xi&=&{2N'_\pm\Delta\gamma v r^2\over\Phi},
\eqe
and a dimensionless quantity $\phat$ that is related to the magnetic pressure in
the cold part of the wind:  
\eqb
\phat&=&{r^2\gamma^2 v {\binfty}^2\over4\pi\Psi}.
\eqe
The continuity equation then simply relates the 
density of cold plasma $\ncold'$ to $\xi$:
\eqb
\xi+{\gamma v r^2\ncold'\over\Phi}&=&1,
\label{continuity3}
\eqe
and the energy equation becomes
\eqb
{\gamma\over \michelmu}+
\left[1+{2-\gheat\over2\left(\gheat-1\right)}\Delta\right]\phat&=&1,
\label{energy3}
\eqe
where the constant $\michelmu$ is the total energy carried by the wind
per unit rest mass ($\times c^2$):
\eqb
\michelmu&=&{\Psi\over m c^2\Phi}\,=\,{L\over 2\dot{N}_\pm m c^2},
\lab{eq:mu_michel}
\eqe
as introduced by \citet{michel69}. 
Physically, $\left(1-\Delta\right)\phat$ is the fraction of the wind 
luminosity carried by magnetic field, $\gamma/\michelmu$ 
is the fraction carried by
the particle rest mass and the remainder, 
$\phat\gheat\Delta/[2(\gheat-1)]$, can be
attributed to the enthalpy flux. 
The constant $\michelmu$ is related to the more commonly used 
$\sigma$ \citep{kennelcoroniti84,kirkduffy99}, defined as the ratio of
Poynting flux to particle-born energy flux by
\eqb
\sigma&=&\left(1-\Delta\right)\phat
\left[{\gamma\over\michelmu}+
\left({\gheat\over\gheat-1}\right)\phat\Delta\right]^{-1},
\label{defsigma}
\eqe
where $\sigma$ is, in general a function of $r$, but is constant in cold 
($\Delta=0$), spherically symmetric flows of constant speed,
in which case $\sigma=\michelmu\phat/\gamma\approx\mu/\gamma$, for 
$\michelmu\gg1$.

The entropy equation is 
\eqb
\lefteqn{
\left({2-\gheat\over\gheat-1}\right){\diff \Delta\over \diff \ln r}
+\left[1+\left({2-\gheat\over\gheat-1}\right)\Delta\right]
{\diff \ln\phat\over \diff \ln r}
}&&
\nonumber\\
&&+\left[{c^2\over\gamma^2 v^2}-\left({2-\gheat\over\gheat-1}\right)\Delta\right]
{\diff \ln\gamma\over \diff \ln r}
+2\Delta\,=\,0,
\label{entropy3}
\eqe
in which, for simplicity, we have assumed $\gheat$ is constant,
and the condition of flux freezing in the cold part of the flow is
\eqb
{\phat v\over
\left(1-\xi\right)^2}&=&{\rm constant}.
\label{freeze3}
\eqe

Thus the conservation equations and the entropy equation 
yield three equations (\ref{energy3}), (\ref{entropy3}) and
(\ref{freeze3}) in the four unknowns $\xi$, $\gamma$, $\phat$ and
$\Delta$. 
\footnote{ The current sheet model adopted by \citet{lyubarskykirk01} 
does not allow cold and hot components of the particle distribution 
to coexist at the same position. This leads to a slightly different 
expression for the density of cold particles:
$\xi+{\gamma v r^2\ncold'(1-\Delta)/\Phi}=1$,
instead of Eq.~(\ref{continuity3}), and, consequently 
for the flux freezing condition:
${\phat\left(1-\Delta\right)^2 v/(1-\xi)^2}={\rm constant}$, 
instead of Eq.~(\ref{freeze3}).}

To close this set, it is necessary to specify the rate at which
dissipation occurs using one of the prescriptions discussed below. 
An additional physical constraint on these can be found by
requiring that the ohmic dissipation term in the entropy equation 
be positive. Performing the perturbation expansion as indicated above
and described in Appendix~\ref{appendixentropy}, this
gives the condition
\eqb
{\diff\over\diff\ln r}
\left[\phat^{1/\gheat}\left(r^2 v/\gamma^2\right)^{(\gheat-1)/\gheat}\Delta
\right]&>&0.
\label{evolutioncondition}
\eqe

\section{Dissipation prescriptions}
\label{reconnection}\lab{sec:reconnection}
In the following, three different prescriptions for the effective,
globally averaged rate of dissipation are introduced. Each
prescription has a specific, physical motivation and corresponds in a
resistive MHD analogy to a certain value for the effective magnetic
diffusivity, as discussed in Sect.~\ref{diffusivity}.
  
\subsection{Slow dissipation}
\label{recipecoroniti}
The current sheet described above can be realized only if
the drift speed required by Eq.~(\ref{driftspeed}) is subluminal. This
gives an absolute lower limit on the thickness of the sheet, or
alternatively, on the fraction of particles it contains. 
The Debye length can be written as:
\eqb
{\ldebye\over a}
&=&{\michelmu r\over\rlight\xi\gamma}\left({\phat v\over\lhat c}\right)^{1/2},
\label{debye2}
\eqe
where we have defined the dimensionless flow luminosity
\eqb
\lhat&=&
\left({\pi^3e^2\Psi\over m^2c^5}\right).
\label{definelhat}
\eqe
Then the condition for a subluminal drift speed, $\beta\le1$, becomes
\eqb
\xi\,\ge\,\xi_{\rm min}
&=&{\michelmu r\over\rlight\gamma}\left({\phat v\over\lhat c}\right)^{1/2}.
\label{coroniti}
\eqe
The treatments of 
both \citet{coroniti90} and \citet{michel94} amount to setting
$\xi=\xi_{\rm min}$.
\citet{lyubarskykirk01} also used this condition. 
A current sheet with particle distributions that are in
equilibrium does not, of course, dissipate magnetic energy. However, 
as the sheet moves outwards with the flow, the parameters describing
it change and it is this slow evolution through a sequence of
quasi-equilibria which implies dissipation. Eq.~(\ref{coroniti})  
leads to a lower limit on the dissipation rate, since a larger
value of $\xi$ implies a thicker sheet (larger $\Delta$) and smaller
fraction of the total energy flux carried by the fields.

\subsection{Fast dissipation}
\label{recipefast}
The maximum dissipation rate is more difficult to estimate. It
follows from the maximum speed at which the current sheet can
expand into the surrounding plasma 
and still remain in the neighbourhood of the Harris
equilibrium. Denoting this speed by $c\vchar$ ($>0$),
the associated rate of evolution of the sheet thickness with radius,
can be written:
\eqb
{\diff\Delta\over\diff\ln r}&=& {2\vchar r c^2\over
\pi\rlight\gamma^2\left(v^2-c^2\vchar^2\right)},
\label{fixedspeed}
\eqe
(see Appendix~\ref{appendixspeed}).

In the outer parts of the sheet, the balance between
magnetic and particle pressure can no longer be maintained
if the bulk velocity exceeds the local speed of the fast magnetosonic
mode $c\vfast$. This suggests the condition
\eqb
\vchar&\le&\vfast\,=\,
\sqrt{B'^2/4\pi\over
\ncold'mc^2+B'^2/4\pi},
\nonumber\\
&=&\left[1+{\gamma\over \michelmu\phat}\left(1-\xi\right)\right]^{-1/2}.
\label{fastspeed}
\eqe
\citet{drenkhahn02} has generalized this prescription,
introducing a parameter $\epsilon=\vchar/\vfast$ and 
arguing that it is typically of the order of unity.

However, in the absence of a component of the magnetic field in the
$\unittheta$ direction, the Harris equilibrium has at its center a neutral
layer. This can only be maintained close to equilibrium if the expansion speed
is small compared to the sound speed in the unmagnetized gas.
Were this 
not the case, the layer would lose its internal pressure support and be crushed by
the external magnetic pressure.
In the
case of a relativistic temperature ($T\gg mc^2$) this leads to the condition
\eqb
\vchar\le {1\over\sqrt{3}}.
\label{soundspeed}
\eqe
In the outer parts of the flow, where the plasma temperature decreases to
below the electron rest mass, condition (\ref{soundspeed}) is
inapplicable and for $T\ll mc^2$, the sound speed is 
$\sqrt{\gheat T/(m c^2)}$.
A full treatment of the dynamics in this regime requires a local evaluation of the 
equation of state, i.e., of the
effective ratio of specific heats $\gheat$. In this paper, we avoid
such complications by taking $\gheat=4/3$ everywhere in the current sheet,
which is appropriate for any isotropic gas in which the particles are
relativistic ($E\approx cp$).
Nevertheless,
we can take approximate account of a more realistic sound speed by setting
\eqb
\vchar\le {\rm Min}\left({1\over\sqrt{3}},\sqrt{T\over mc^2}\right)
\enspace.
\label{realisticsound}
\eqe
In flows in which the Poynting flux dominates, this restriction 
leads to a much more stringent upper limit on the dissipation rate 
than does Eq.~(\ref{fastspeed}), since $\vfast\approx 1$, for
$\sigma\gg 1$,
and the denominator on the right-hand side of Eq.~(\ref{fixedspeed})
becomes small.
The limit can be relaxed slightly if an
additional component of the magnetic field 
(in the $\unittheta$ direction) is present in the sheet.
In this case the local speed of the fast mode would be more appropriate
than the speed of sound in an unmagnetized plasma. 
However, 
this makes only a minor difference, 
provided the plasma at the center of the sheet
is not too far from equipartition.

\subsection{Dissipation limited by the tearing mode}
\label{recipetearing}
The Harris equilibrium is well-known to be unstable 
to the excitation of a variety of waves in the non-relativistic case
--- see, for example, 
\citet{daughton99,biskamp00}. It is thought that the 
current sheet first filaments because of the growth of the tearing mode,
producing magnetic islands separated by neutral lines. These then lead to
magnetic reconnection and its associated dissipation \citep{melrose86}.
This scenario has largely been developed in the context of the geomagnetic 
tail of the Earth. Particle-in-cell simulations appear to 
confirm its validity \citep{zhuwinglee96,pritchettetal96}, although the
3-dimensional picture is more complicated 
\citep{buechnerkuska99,hesseetal01}

In the relativistic case 
the linear growth rate 
of the tearing instability has been calculated 
by \citet{zelenyikrasnoselskikh79}. For $T\gg mc^2$ they find 
for the most unstable mode,
which has a wavelength approximately equal to the sheet thickness,
\eqb
\gtear&=&\beta^{3/2}c/a,
\nonumber\\&=&\ldebye^{3/2}c/a^{5/2}.
\label{teargrowth}
\eqe
Thus, for a given number of particles in the sheet, and 
fixed external pressure (implying $\xi$, $\phat$ and $\beta$ 
constant), the growth rate is faster for thinner
sheets. 
\citet{lyubarsky96}, in his model of the high energy emission from the
Crab pulsar, suggested that the speed $\vchar$ of expansion of the sheet
could be self-regulating being governed by the fastest growing
mode of the relativistic collisionless tearing mode instability:
\eqb
\vchar&=&a\gtear/c,
\eqe
or, in terms of the flow variables,
\eqb
{\diff\left(\gamma\Delta\right)\over\diff\ln r}&=&{2r c\over\pi\rlight\gamma v}
\left({\xi_{\rm min}\over\xi}\right)^{3/2}.
\label{recipejuri}
\eqe

\section{Solutions of the flow equations}
\label{results}
In this section we estimate both the length scale on
which dissipation converts an initially Poynting-dominated flow into 
one in rough equipartition, as well as the rate at which the flow accelerates,
i.e., $\gamma(r)$ during this process. To simplify the results, we specialize
to the case of a relativistic gas: $\gheat=4/3$.

At an initial radius, assume the flow contains a thin 
sheet $\Delta\ll1$ containing relatively few particles  $\xi\ll1$, is
Poynting flux dominated with $\gamma/\michelmu\ll1$ and already super magnetosonic
$\gamma >\michelmu^{1/3}$. As the radius increases, the flow accelerates,
but, as long as the above ordering is maintained, we may set $\phat\approx1$
and $v\approx c$. 
In this range, the flux freezing equation (\ref{freeze3}) 
reduces to the approximate relation
\eqb
{\gamma\over \michelmu}-2\xi+\Delta
&=&{\rm constant},
\label{freezeapprox}\lab{eq:freezeapprox}
\eqe
and the entropy equation is 
\eqb
{\diff\over\diff\ln r}\Delta-\left(2\Delta+{\gamma\over\michelmu}\right)
{\diff\over\diff\ln r}\ln\gamma + 2 \Delta&=&0.
\label{approxentropy}
\eqe

When these two equations are complemented by one of the dissipation
prescriptions [using, for fast dissipation, Eq.~(\ref{soundspeed})], 
it is straightforward to obtain asymptotic solutions
for the flow for large radius. These take the form
\eqb
\Delta&\propto& r^q,
\label{asymptoticdelta}\\
\xi&=&{\Delta\over q},
\label{asymptoticxi}\\
{\gamma\over\mu}&=&{\Delta(2-q)\over q},
\label{asymptoticgamma}
\eqe
[or, in the case of the LK sheet,
$\xi=\Delta(1+q)/q$ instead of Eq.~(\ref{asymptoticxi})].
Shortly before dissipation is complete, at the point where $\phat$
decreases, $\Delta\approx\xi\approx1$ and $\gamma\approx\michelmu$, 
the solutions lose their validity. For each
dissipation prescription, the asymptotic solutions yield reasonably accurate
estimates of the radius at which this occurs, as well as the dependence of 
the flow variables on radius up to this point. An estimate of the smallest
radius at which the asymptotic solutions are valid is obtained by locating the
magnetosonic point. These results are summarized in
Table~\ref{tableasymptotics}. 
The radius at which dissipation is complete can be estimated by
inserting into the asymptotic solution either $\Delta=1$, $\xi=1$, or
$\gamma=\michelmu$ and the numerical integrations indicate that the latter
is somewhat more accurate.
Table~\ref{tableasymptotics} gives this radius with a numerically
estimated coefficient to order of magnitude accuracy. 
For slow dissipation this asymptotic solution was originally found by 
\citet{lyubarskykirk01}; for fast dissipation, the scaling with radius agrees with
that given by \citet{drenkhahn02}. 

\begin{deluxetable}{p{3.2cm}|cccc}
\tabletypesize{\scriptsize}
\tablecaption{\label{tableasymptotics}
Asymptotic flow properties for different dissipation prescriptions. 
Given are the index $q$ in Eqs.~(\ref{asymptoticdelta}--\ref{asymptoticxi}), 
the approximate radius $r_{\rm max}$ at which dissipation is complete,
the approximate radius $r_{\rm min}$ at which the solution crosses the 
fast magnetosonic point and the exact (asymptotic) expression for
$\Delta$. The light-cylinder radius is denoted by $\rlight$, $\Delta$
is the thickness in units of the pattern half-wavelength, $\michelmu$
is the total energy carried by the wind per unit rest mass ($\times
c^2$), $\lhat$ is the dimensionless flow luminosity defined in
Eq.~(\ref{definelhat}), and $\vchar$ is the sheet expansion speed in
the comoving frame, in units of $c$.
}
\tablehead{
& \colhead{$q$}    & \colhead{$r_{\rm max}/\rlight$}   &
\colhead{$r_{\rm min}/\rlight$}   &   \colhead{$\Delta\left(r/\rlight\right)^{-q}$}}        
\startdata
Slow dissipation\newline 
(Eq.~\ref{coroniti})
         &  1/2        &  $\lhat^{1/2}$                           & $\michelmu^{-4/3}\lhat^{1/2}$               
                        &${1\over\sqrt{6}}\lhat^{-1/4}$    \\
&&&&\\

Tearing mode limited\newline 
dissipation (Eq.~\ref{recipejuri})
         &  5/12       & $\michelmu^{4/5}\lhat^{3/10}$            & $\michelmu^{-4/5}\lhat^{3/10}$         
            &$\left({5^{8}\over19^7 12\pi^{2}}\right)^{1/12}\michelmu^{-1/3}\lhat^{-1/8}$ \\
&&&&\\

Fast dissipation\newline 
(Eqs.~\ref{fixedspeed} \& \ref{soundspeed})
         &      1/3    & $0.1\michelmu^2(1-\vchar^2)/\vchar$                            & $0.1(1-\vchar^2)/\vchar$  
            &$\lft({6\vchar\over 25\pi(1-\vchar^2)}\rgt)^{1/3}\michelmu^{-2/3}$\\
\enddata
\end{deluxetable}

\begin{figure}
\epsscale{.65}
\plotone{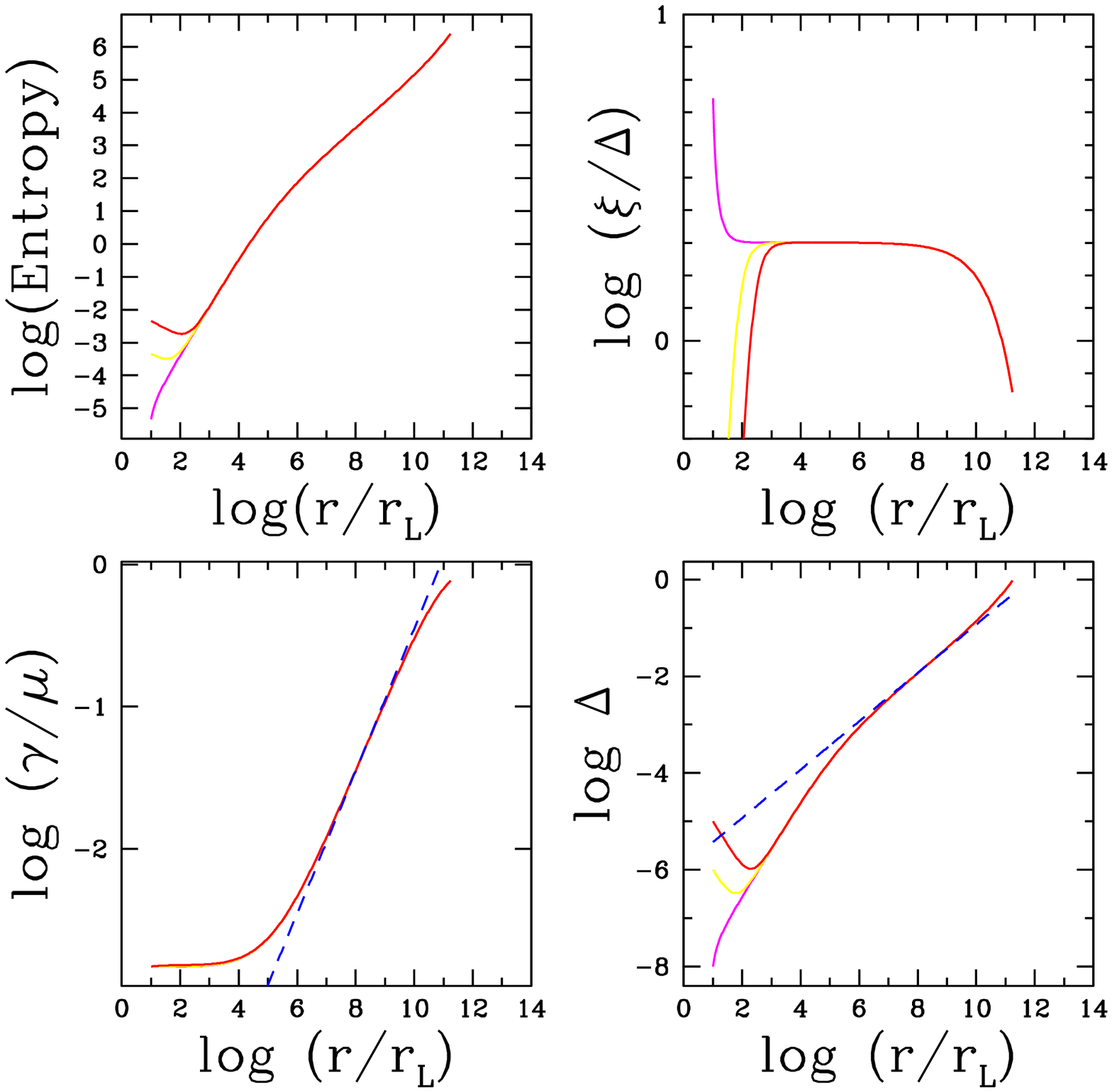}
\plotone{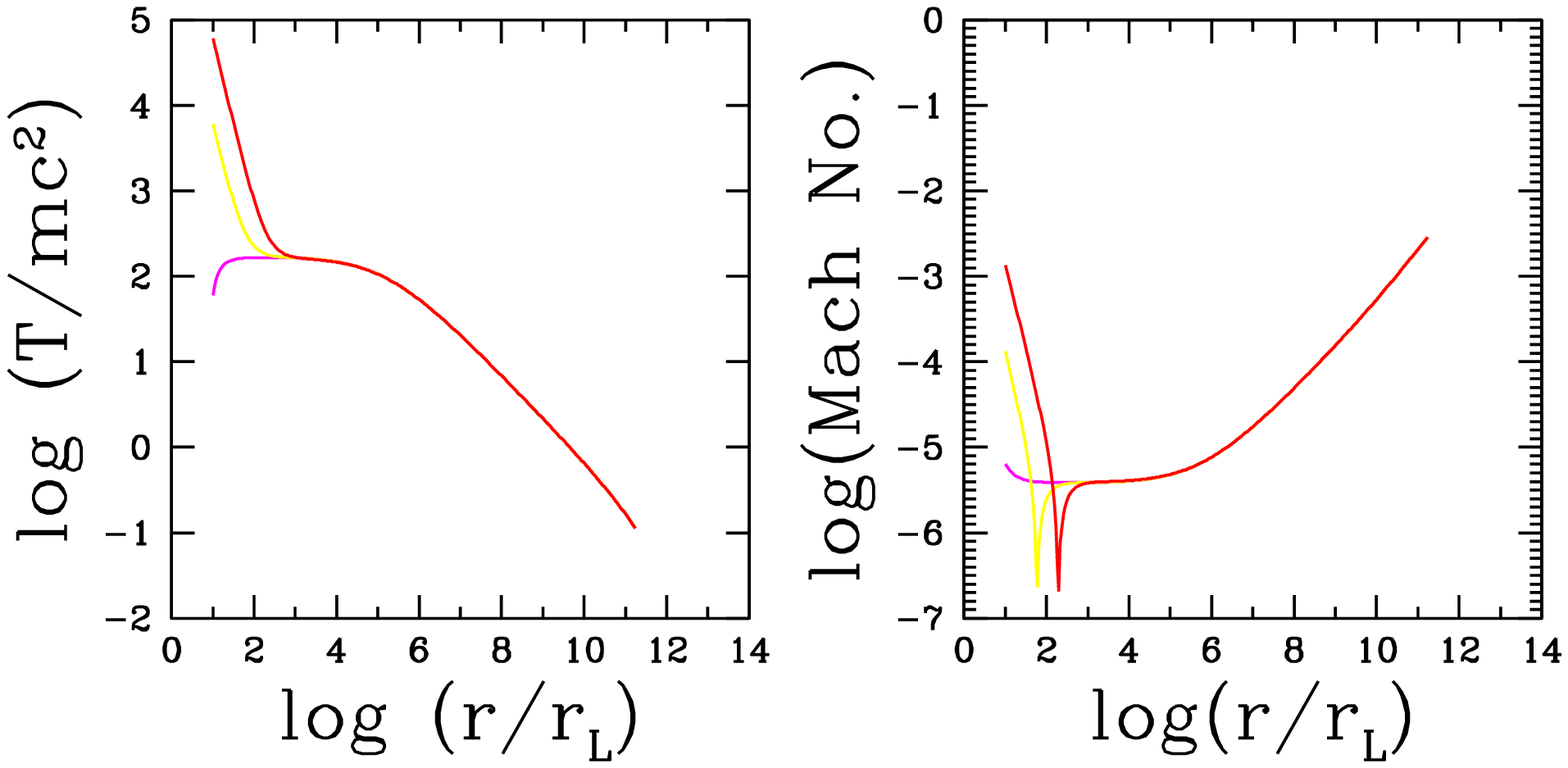}
\caption{The flow variables as functions of radius for
slow dissipation and parameters $\lhat=1.5\times10^{22}$, $\omegawind = 4\pi$, $\michelmu=
2\times10^4$, thought appropriate for the Crab pulsar. Displayed are the
average entropy [see Eq.~(\ref{evolutioncondition})], 
the bulk Lorentz factor, the fraction of a wavelength occupied by the 
hot sheet $\Delta$, the temperature and the Mach number of the
expansion speed of the sheet. The top right panel shows the ratio $\xi/\Delta$
(which gives the ratio between the density of hot particles in the sheet,
and the wavelength-averaged density of both hot and cold particles).
Three numerical solutions are 
shown as solid lines for different initial conditions on $\Delta$. 
The asymptotic solutions for $\gamma$ and $\Delta$ are shown as dashed lines.}
\label{numericalslow}
\end{figure}

\begin{figure}
\epsscale{.65}
\plotone{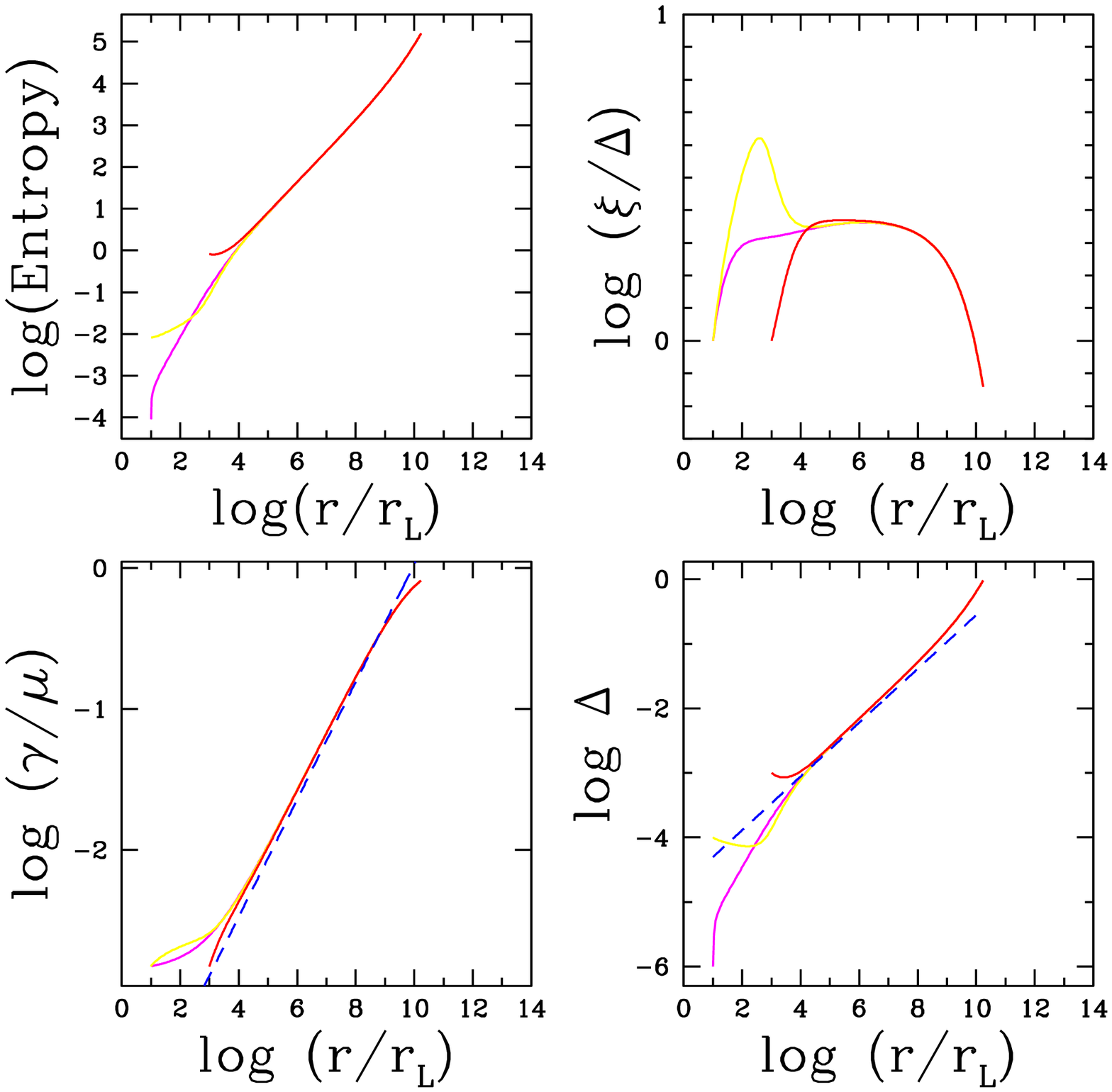}
\plotone{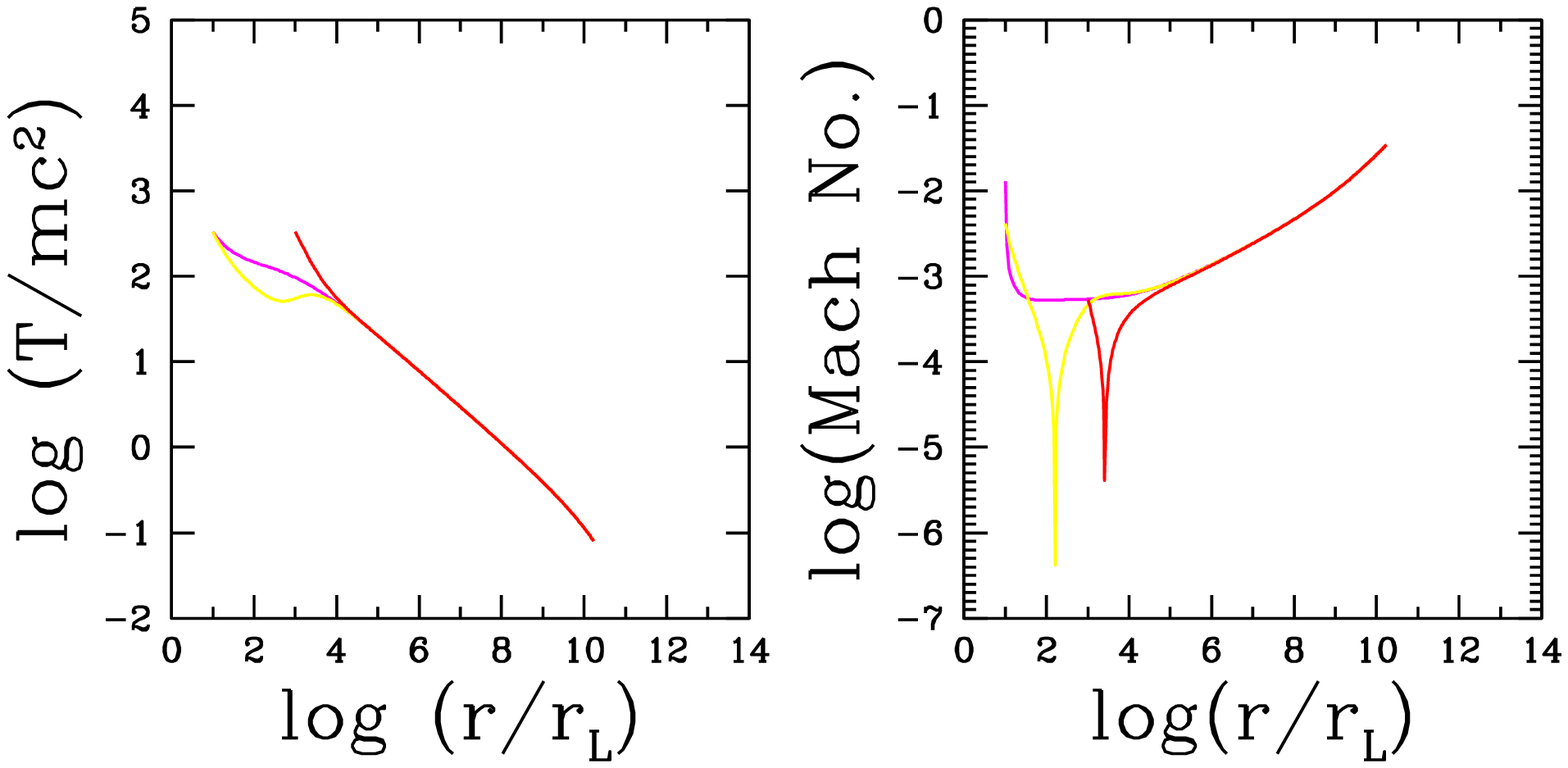}
\caption{The flow variables as functions of radius for
tearing mode limited dissipation (parameters as in Fig.~\ref{numericalslow}).
Three solutions are shown as solid lines for initial conditions
$\Delta=10^{-6}$ and $10^{-4}$ starting at $r=10\rlight$, and for
$\Delta=10^{-3}$ starting at $r=1000\rlight$. The asymptotic solution is shown
as a dashed line.}
\label{numericaltear}
\end{figure}

\begin{figure}
\epsscale{0.65}
\plotone{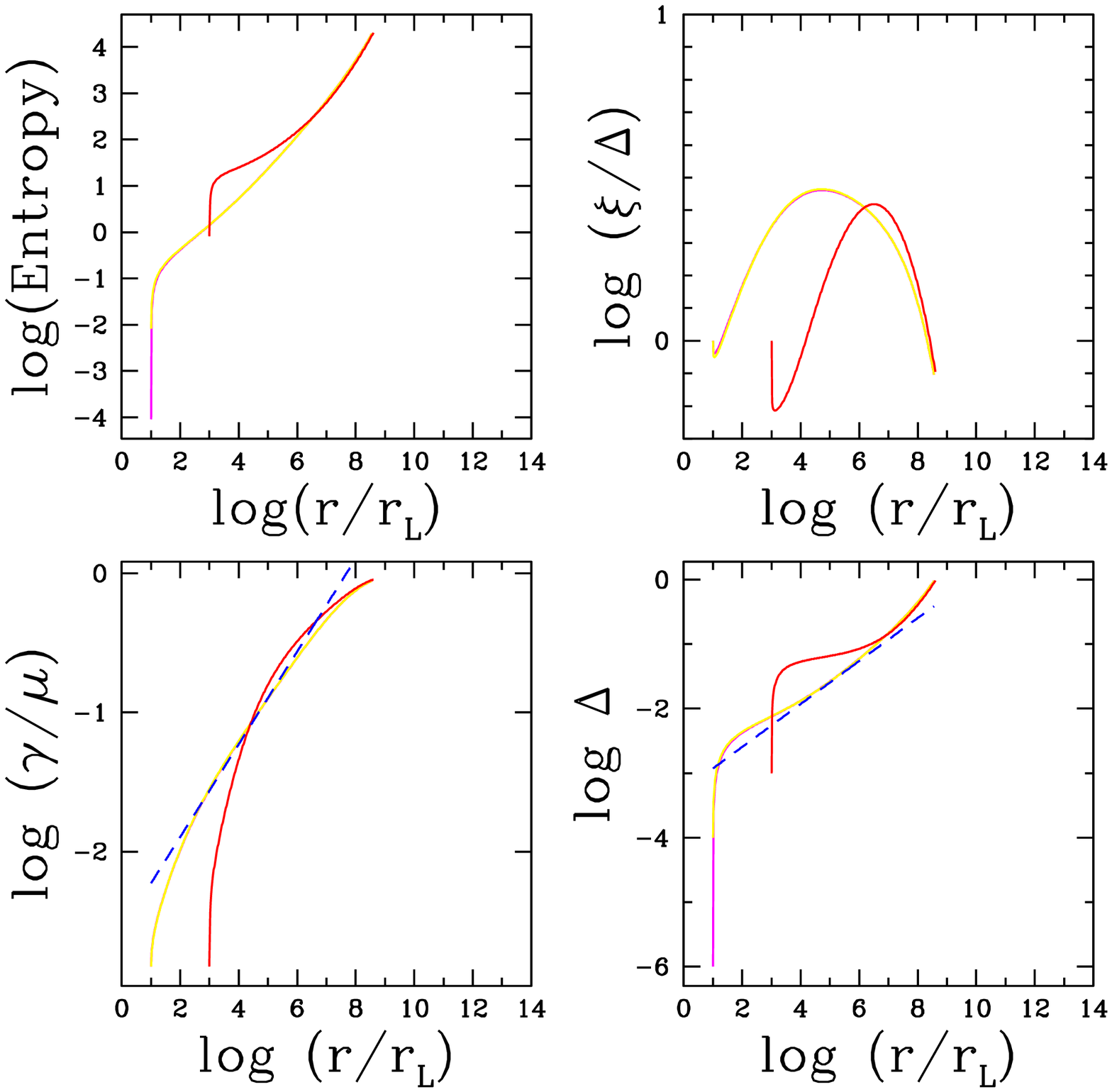}
\plotone{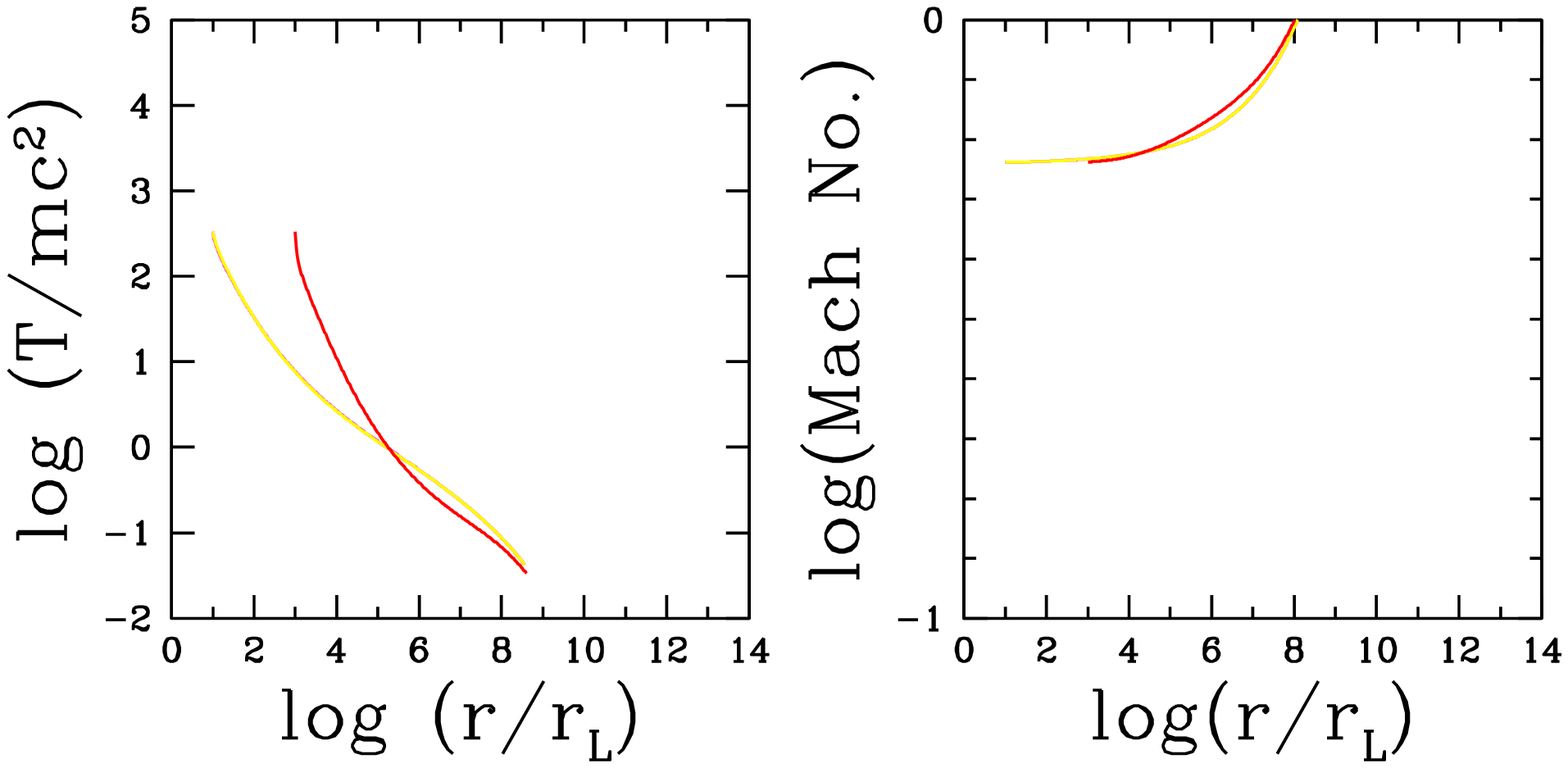}
\caption{The flow variables as functions of radius for
fast dissipation according to Eqs.~(\ref{fixedspeed}) and
(\ref{soundspeed}), 
for the same parameters and initial conditions as in
Fig.~\ref{numericaltear}.
The solutions with initial values of $\Delta=10^{-6}$ and
$\Delta=10^{-4}$ coincide for almost all radii.}
\label{numericalfast}
\end{figure}

Figure~\ref{numericalslow} compares the results of a numerical integration of
the full equation set, Eqs.~(\ref{energy3}), (\ref{entropy3}) and (\ref{freeze3}), 
together with the prescription for slow dissipation
in which $\xi$ is required to equal $\xi_{\rm min}$, defined in Eq.~(\ref{coroniti}). 
The parameters chosen in this 
and the following three figures are those implied by the discussion of the Crab pulsar 
in Sect.~\ref{discussion}:
$\lhat=1.5\times10^{22}$ and $\michelmu=2\times10^{4}$. Six quantities are
plotted against radius: 
\begin{enumerate}
\item
the quantity defined in 
Eq.~(\ref{evolutioncondition}) which is proportional to the 
entropy averaged over a wavelength
\item
the ratio $\xi/\Delta$, that is the ratio of the density of hot
particles to the wavelength-averaged particle density
\item
the Lorentz factor $\gamma$ (in units of its maximum value $\michelmu$) 
\item
the fraction $\Delta$ of a wavelength occupied by hot plasma 
\item
the temperature 
\item
the absolute value of the Mach number 
(with respect to the fast magnetosonic speed in the cold part of the
wind) with which the
sheet expands. 
\end{enumerate}

\begin{figure}
\epsscale{0.65}
\plotone{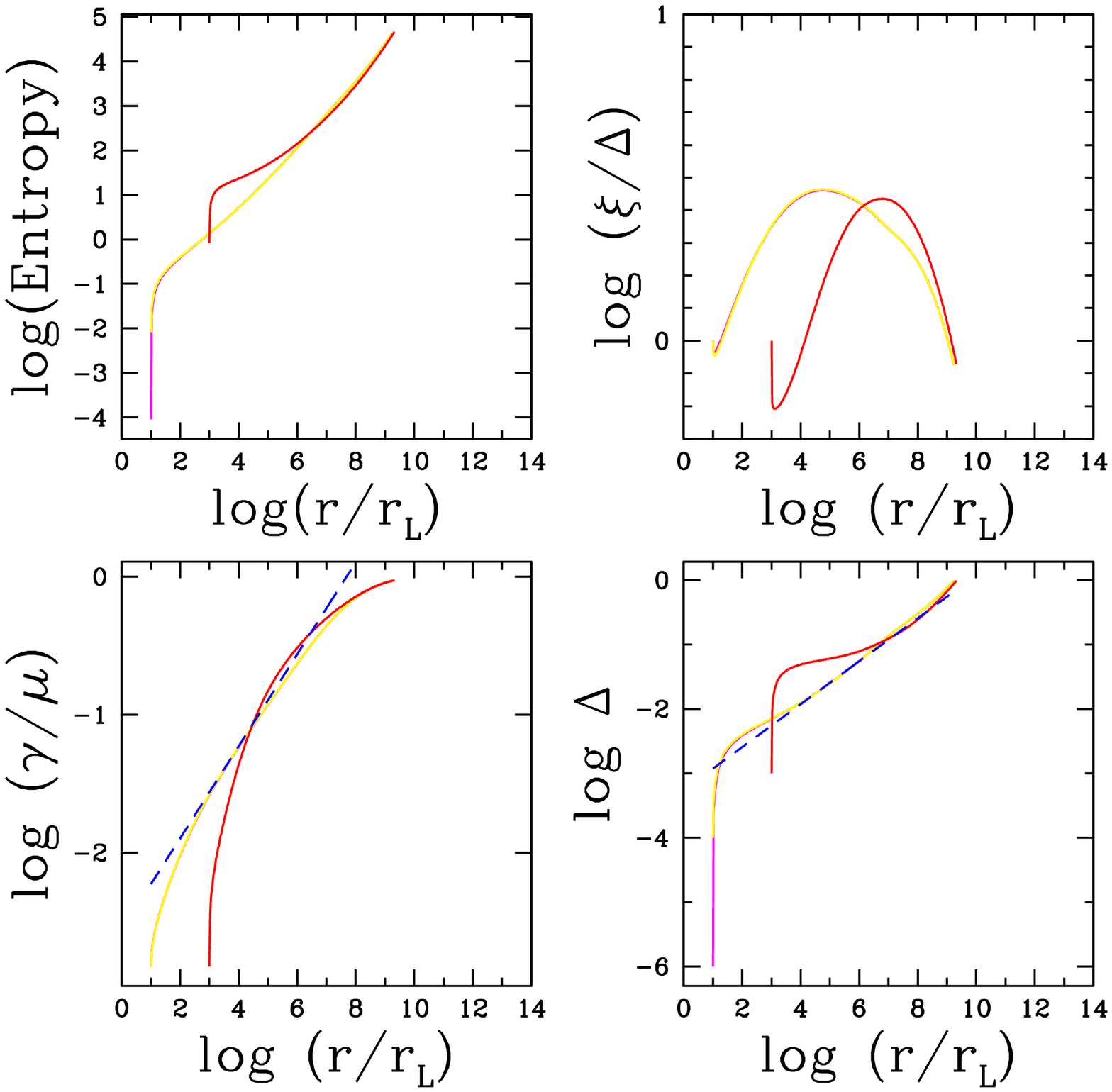}
\plotone{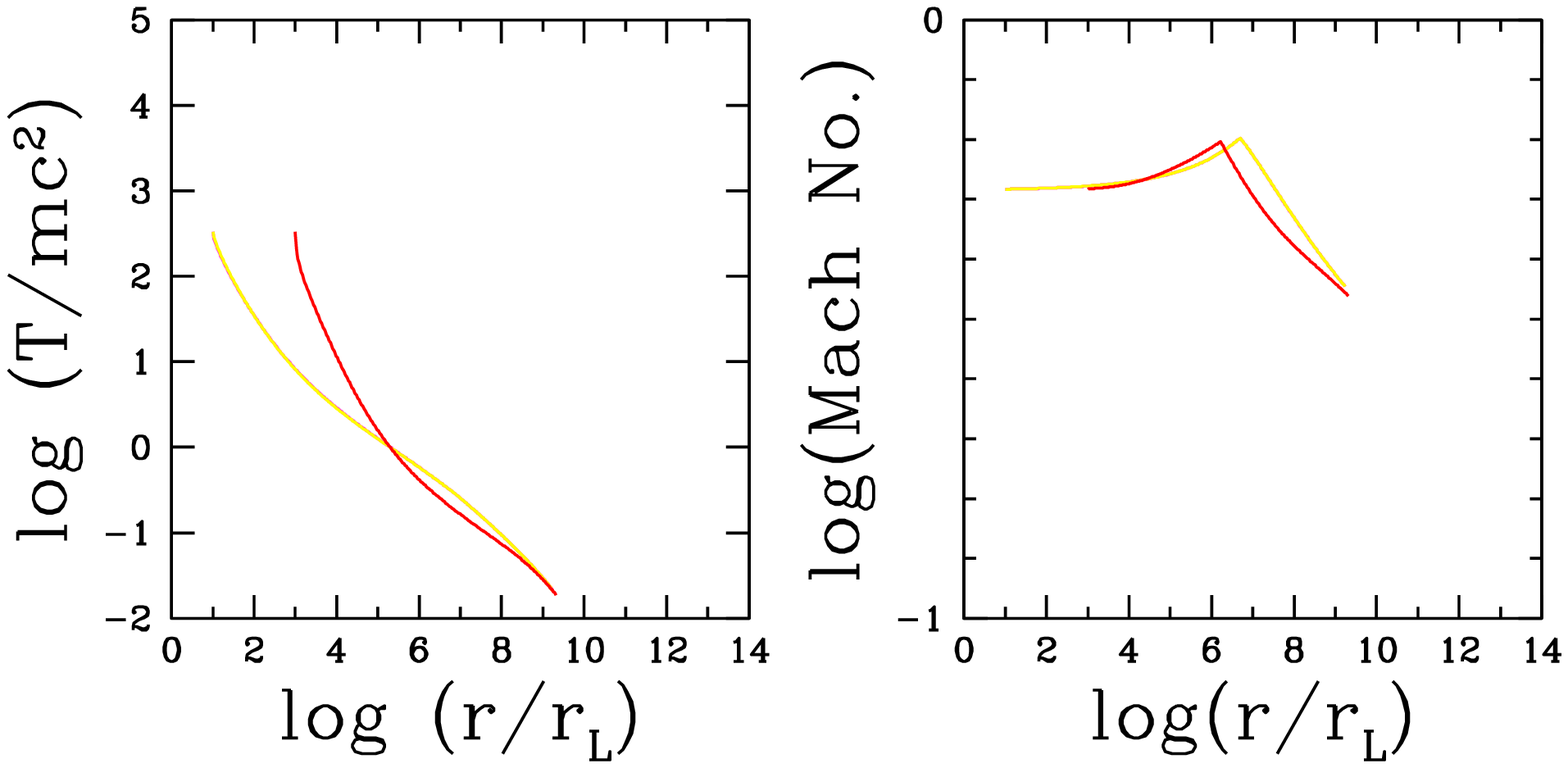}
\caption{The flow variables as functions of radius for
fast dissipation using a realistic estimate of the sound speed
according to Eq.~(\ref{realisticsound}), and for the same parameters
and initial conditions as in Fig.~\ref{numericalfast}. The solutions
with initial conditions $\Delta=10^{-6}$ and $\Delta=10^{-4}$ coincide
for almost all radii.}
\label{numericalrealistic}
\end{figure}

Solutions are shown for three different sets of initial conditions with 
$\Delta=10^{-8}$, $10^{-6}$ and $10^{-5}$,  
all starting at the point $r=10\rlight$ with a slightly supermagnetosonic Lorentz 
factor $\gamma=1.1\mu^{1/3}\approx33$.
All converge rapidly to the asymptotic solution given in Table~\ref{tableasymptotics} and 
Eqs.~(\ref{asymptoticdelta}-\ref{asymptoticgamma}),
which, in the case of $\gamma$ and $\Delta$, is plotted as a dashed line. Deviations from 
this solution are small and appear only close to the 
maximum radius, where dissipation is complete. Using this prescription for dissipation, 
it is possible to choose an arbitrarily small initial value of $\Delta$ --- 
the results do not differ appreciably from the case $\Delta=10^{-8}$. 
However, these solutions adjust rapidly in the initial phase
and cannot strictly be described by the 
perturbation approach underlying our method, which assumes all variations to be on a length scale 
equal to or larger than the local radius.
For the larger initial values of $\Delta$, the region 
occupied by hot plasma shrinks as the flow moves outwards. This is accompanied by a decrease 
in the averaged entropy, indicating that 
the prescription is unphysical. Thus, we cannot be sure that initial states with such a large 
value of $\Delta$ will
approach the asymptotic solution, since a different dissipation prescription with increased 
entropy generation is required at least initially. 
Different values of the starting point in radius yield qualitatively similar solutions.

Figure~\ref{numericaltear} shows results for the case of tearing mode limited dissipation, 
Eq.~(\ref{recipejuri}). 
Three different sets of 
initial conditions were chosen: $\Delta=10^{-6}$ and $10^{-4}$ starting at $r=10\rlight$ 
and $\Delta=10^{-3}$ starting at $r=1000\rlight$. 
Because this prescription 
takes the form of a first order differential equation, an additional initial condition on $\xi$ 
is needed --- we choose $\xi=\Delta$.
Once again, the asymptotic solution provides a reasonable approximation at large radius. In this case,
however, the entropy increases in all solutions, despite the fact that the region occupied by hot plasma shrinks slightly.
Thus, this prescription provides a physically consistent description of the approach of the flow to the asymptotic solution even for
quite large initial values of $\Delta$. However, the initial density contrast $\xi/\Delta$ 
between the hot and cold parts of the flow has to be chosen with 
with some care in order to avoid rapidly varying transients. At
$r>10^8\rlight$, the plasma temperature is nonrelativistic, $T<mc^2$
and the expression used for the growth rate of the tearing mode
instability loses  validity.

The case of fast dissipation is shown in Fig.~\ref{numericalfast},
where we have used the relativistic sound speed (\ref{soundspeed}) as
an upper limit on the expansion rate of the sheet. Here again, the dissipation prescription takes
the form of an ordinary differential equation, so that initial conditions on both $\Delta$ and $\xi$ are required.
These were chosen to be the same as those of Fig.~\ref{numericaltear}. The solutions again converge to the asymptotic 
expressions, and the entropy rises throughout. However, the initial transients fluctuate more strongly here than
in the case of the tearing mode limited prescription. The result of
the more realistic estimate of the sound speed given by
Eq.~(\ref{realisticsound}) is shown in
Fig.~\ref{numericalrealistic}, for the same parameters and initial
conditions as used in Fig.~\ref{numericalfast}. Differences between
these two sets of solutions appear only in the outer parts of the
flow, where the temperature of the plasma becomes nonrelativistic. The
Mach number drops sharply at this point, and the 
dissipation rate is reduced. Independent of the initial conditions,
the solutions terminate at $r\approx2\times10^9\rlight$.
No asymptotic solutions are available 
to describe this outer phase of the flow. Nevertheless, the estimates of the
radius at which dissipation is complete obtained using the asymptotic
solutions
for fast reconnection given in Table~\ref{tableasymptotics} remain accurate to within 
roughly an order of magnitude.

\section{Comparison with dissipation by Bohm diffusion}
\label{diffusivity}
For each dissipation prescription, the effective relaxation time of
the plasma can be estimated by appealing to an MHD picture: the
magnetic diffusivity $\eta$ is related to the sheet thickness and the
dynamical evolution time in the comoving frame $\tdyn$ by
\eqb
\eta&\approx&{a^2\over \tdyn}
\eqe
and is given in terms of the effective collision time $\trelax$ by
\eqb
\eta&=&{\ldebye^2\over\trelax},
\eqe
where we have used $\ldebye$ instead of the Larmor radius, as 
appropriate in a relativistic plasma (in the current model the two
differ by a constant factor of the order of unity). 
Combining these two equations
gives
\eqb
\trelax&\approx&\left({\ldebye\over a}\right)^2\tdyn
\label{relax2}
\eqe
and we see that the collision time is shorter than the dynamical time provided
$a>\ldebye$, or equivalently, $\xi>\xi_{\rm min}$.

As can be seen from Eq.~(\ref{asymptoticdelta}), the asymptotic
solutions presented in Sect.~\ref{results} correspond to a dynamical
timescale for the current sheet in the comoving frame given, to within
a factor of the order of unity, by 
\eqb
\tdyn&=&{r\over\gamma c}.
\label{relax}
\eqe
Thus, it is straightforward to compare the effective collision time
implied by the dissipation prescription $\trelax$ with the relaxation
time $\tbohm\approx\ldebye/c$ corresponding to 
\lq\lq Bohm\rq\rq\ diffusivity in a relativistic plasma.
Using Eqs.~(\ref{relax2}) and (\ref{debye2}), we have
\eqb
{\trelax\over\tbohm}&\approx&\left({\ldebye\over
    a}\right)\left({r\over\rlight \Delta\gamma^2}\right)
,\nonumber\\
&=&
{\michelmu r^2\over\xi\gamma^3\rlight^2\Delta}\left({\phat v\over \lhat
  c}\right)^{1/2}.
\eqe
At the outer radius, this ratio takes on its smallest value for slow
dissipation and for dissipation limited by the tearing mode:
\eqb
{\trelax\over\tbohm}&\gtrsim&
\left\lbrace
\begin{array}{ll}
\left({\lhat^{1/4}/\michelmu}\right)^{2} &\textrm{(slow)}\\
\left({\lhat^{1/4}/\michelmu}\right)^{2/5}&\textrm{(tearing mode)}
\label{slowbohm}
\end{array}
\right.
\eqe
and its largest value for fast dissipation:
\eqb
{\trelax\over\tbohm}&\lesssim&
\left({\lhat^{1/4}\over\michelmu}\right)^{-2}\ \textrm{(fast).}
\label{fastbohm}
\eqe

\section{Discussion}
\label{discussion}

For the acceleration of the flow to proceed as we have described, the actual
dissipation rate must lie above that of \lq slow\rq\ dissipation and
below that of \rq fast\rq\ dissipation. To a reasonable approximation, these
conditions can be imposed simply by inspecting the appropriate values of
$r_{\rm max}$ and $r_{\rm min}$, as given in Table~\ref{tableasymptotics}. It
is immediately clear that no prescription can fulfil these requirements
unless $\michelmu$ is less than a certain maximum value. Adopting
$\vchar=1/\sqrt{3}$, we find
\eqb
\michelmu&<&3\lhat^{1/4}
\enspace.
\label{absolute}
\eqe
Physically, restriction (\ref{absolute}) 
arises because at higher values of $\michelmu$ the sheet would have to expand
supersonically in order to accumulate enough particles to maintain the
required current. This restriction was pointed out by LK, who speculated that,
when it is
violated, dissipation ceases before the available magnetic energy is exhausted. 
Whether this is the ultimate fate of such a wind is not completely
clear. It is certainly true that our treatment using the Harris sheet
or a model with hot and cold fluids fails in this regime, and it seems
probable that no equilibrium can then exist in which the electric
field vanishes in the comoving frame. Perhaps at this point it is essential to allow
the conversion of the entropy wave into an electromagnetic mode, along the
lines suggested by \cite{melatos02}.

If Eq.~(\ref{absolute}) is fulfilled, Eqs.~(\ref{slowbohm}) and (\ref{fastbohm}) 
show that our solutions imply slower dissipation 
than the Bohm value
at all radii for the \lq slow\rq\ and \lq tearing-mode\rq\ cases, 
but faster-than-Bohm dissipation in the \lq fast\rq\ case. In standard
treatments of reconnection, this value is often taken as an
upper limit for the diffusivity
\citep{somovtitov85,priestforbes00,lyutikovuzdensky02}. A higher value
is justified, however, if the effective internal diffusion scale in
the current sheet is given by a turbulent mixing length, rather than
the characteristic Larmor radius of the particles. A turbulent sheet
structure arises if the sheet filaments into a large number of
reconnection centers when field dissipation begins. In the model
presented here, it is assumed that the global sheet evolution can
nevertheless be described within an MHD framework, using an effective
average value for the dissipation rate.

As the wind approaches the point where dissipation is complete, the
plasma temperature drops, as can be seen from Figs.~\ref{numericalslow}--\ref{numericalrealistic},
and the approximation of a relativistic gas
fails. One effect of this is to invalidate our choice of a fixed ratio
of specific heats $\gheat=4/3$ for the gas dynamics. However, more
importantly, the dissipation prescriptions we have used are also
affected. The Harris equilibrium gives a general expression for the
drift speed of particles in the sheet, valid in both the relativistic and 
nonrelativistic cases, so that the prescription for slow dissipation
remains a lower limit. The upper limit given by the
limiting the speed of expansion of the sheet to that of sound 
in an unmagnetized gas can also be implemented in the nonrelativistic
case, as discussed in Sect.~\ref{recipefast} and shown in Fig.~\ref{numericalrealistic}.
However, in the case of
dissipation
limited by the growth rate of the tearing mode, the physical basis of 
our estimates is no longer valid once the plasma temperature drops
below the particle rest mass. An improved treatment, using 
an expression for the growth rate valid in the transrelativistic and
nonrelativistic regimes is possible, but our estimates are in any case
very rough, and the added complications would
not appear to be justified.

However, given that sufficient particles are present in the wind to satisfy condition
(\ref{absolute}), and neglecting the modification of the 
tearing mode growth rate for nonrelativistic plasma, the prescription for tearing mode limited
dissipation automatically lies between the limiting dissipation rates and can
therefore be considered as a physically realistic possibility. 
For Crab pulsar parameters, $\lhat=1.5\times10^{22}$ and condition
(\ref{absolute}) gives $\michelmu < 10^6$, for $\vchar=1/\sqrt{3}$.
Estimates of the $\michelmu$ parameter for the Crab, based on
modelling the X-ray and gamma-ray emission
\citep{hibschmanarons01a},
exceed this value. In this case the solution obtained by LK for slow dissipation
loses its validity before dissipation is complete. Nevertheless, the range of
validity extends from close to the pulsar to well beyond the radius at which a
termination shock is indicated by observations. X-ray observations 
\citep{weisskopfetal00} place this at a radius of $0.14\,$pc or
$r=2.7\times10^{9}\rlight$ in the equatorial plane. Slow dissipation implies 
steady acceleration of the wind to beyond this radius, without significant
dissipation of the Poynting flux.

Recently, however, two different lines of investigation have suggested that
the 
conventional approach seriously underestimates the rate at which pairs are injected by
the Crab pulsar into its surrounding Nebula. \citet{gallantetal02} have
proposed a model that accounts for the synchrotron spectrum of the Nebula from the radio to hard
gamma-ray bands. In particular, the large number of radio-emitting electrons, which could not
be explained in the model of \cite{kennelcoroniti84}, arise naturally from a
relatively low speed, high density pulsar wind. The estimated 
injection rate is $\dot{N}_\pm\approx 3\times10^{40}\,{\rm s}^{-1}$, 
implying $\michelmu\approx2\times10^4$.
\citet{lyubarskyeichler01},
who analysed the dynamics of the X-ray jet, also conclude that 
a relatively low value of $\sigma$ at the base of the flow is
required. Because of the 
difficulty of collimating a very low density wind, 
they suggest that on the equator $\sigma<10^2$. 
For a transonic wind, this implies $\michelmu=10^3$.  

Such low values of $\michelmu$ and the corresponding high pair densities, 
have important implications for dissipation in the pulsar
wind. Firstly, independent of the actual mechanism, the
asymptotic Lorentz factor of the wind after dissipation equals
$\michelmu$. This fits in well with the spectral modelling of
\citet{gallantetal02},
who correlate this value with the break at $10^{16}\,$Hz in the UV spectrum of
the Crab Nebula.
Secondly, although the radius at which dissipation is complete is independent
of $\michelmu$ in the
\lq slow\rq\ dissipation prescription of \citet{coroniti90}, \citet{michel94}
and LK, this is not true of the faster tearing mode limited case. 
From Table~\ref{tableasymptotics} it is evident that for $\michelmu=2\times10^4$,
dissipation by the tearing mode limited mechanism is still
not complete within the termination shock. However, a faster mechanism 
could provide the required dissipation in this case. 
The lower limit on the dissipation radius implied by the estimates in 
Table~\ref{tableasymptotics} is $5\times10^7\rlight$ and
Fig.~\ref{numericalrealistic} shows that when this limit is modified
to take account of the nonrelativistic sound speed, the solution
dissipates the magnetic field completely at $r=2\times10^9\rlight$, i.e.,
within the termination shock.
Adopting the value $\michelmu=10^3$, or lower, reduces the dissipation radius still
further, giving, for the tearing mode limited prescription,
$r=10^9\rlight$. The minimum distance required to dissipate the
magnetic field into relativistic plasma according to 
Table~\ref{tableasymptotics} is $r>1.5\times10^5\rlight$, and
a numerical integration using the more general limit implied by 
Eq.~(\ref{realisticsound}) gives $r>5\times10^6\rlight$.

It is interesting to note that the appearance of a dissipative wind may differ
between these two cases. \citet{kirkskjaeraasengallant02} 
derived the criterion
$\gamma^2>r/\rlight$ for the radiation emitted at a given radius in a striped
wind to appear pulsed to the observer. 
Accordingly, for fast dissipation, radiation emitted by the hot
current sheet at all positions within the flow should appear
pulsed. On the other hand, the situation for dissipation limited by
the tearing mode growth rate is more complicated: radiation from 
close to the maximum radius should not appear pulsed. However, nearer to
the star the Lorentz factor of the flow is sufficient to ensure a
pulsed appearance. The magnetic field in the flow decreases
with increasing radius, so that the efficiency of
conversion of heat into synchrotron radiation should be greater at
smaller $r$. Consequently, pulsed radiation can be expected also in 
the tearing mode limited case.

\section{Conclusions}
\label{conclusions}
We model the dissipation of an oscillating, toroidal magnetic field
component in a relativistic MHD outflow by 
considering the radial evolution of the entropy wave associated with this pattern. 
Because of magnetic reconnection, or, more precisely, annihilation,  
plasma in the current sheets separating regions of oppositely
directed field is heated. 
We assign an energy density and pressure to the sheets and relate 
them by the equation of state of a relativistic gas.
Upon expanding, the sheet performs work, which
causes the outflow to accelerate.

Despite recent progress \citep{zenitanihoshino01,lyutikovuzdensky02}
the dissipation rate in a relativistically hot pair plasma embedded in
stripes of cold, magnetized  plasma 
cannot be written down
a priori. Thus,
we consider three different prescriptions, each with
a specific physical motivation. 
A lower limit on the dissipation rate is given by the prescription 
used by \cite{coroniti90}, 
\cite{michel94} and
LK, in which the characteristic drift speed in the current
sheet is equal to the speed of light (to within a factor of order unity). 
We argue that an upper limit on the dissipation rate is set by  
allowing the 
sheet to grow at its internal sound speed (in the comoving frame).  
When consistent with these limits, a physically plausible 
estimate of the actual rate is given by assuming that expansion of the
current sheet is
limited by the growth rate of the relativistic collisionless tearing
mode. 
Comparing our prescriptions to a collisional MHD
model, we find that the required dissipation is slower than the Bohm
value when dissipation is limited by the tearing mode, but faster when the
sheet grows at its internal sound speed. However, the complex flow
pattern which will presumably be built up in the dissipation region 
make it difficult to draw conclusions from this analogy.

Analytic, asymptotic solutions, valid for relativistic plasma
temperatures, 
are found for each 
prescription. These agree with and extend those presented by LK for slow
dissipation, with $\gamma\propto r^{1/2}$, and by 
\cite{drenkhahn02} for fast dissipation, with $\gamma\propto r^{1/3}$. For tearing-limited
reconnection we find $\gamma\propto r^{5/12}$. 
Numerical integration of the full equation set
confirms these solutions and demonstrates that
they are valid until
dissipation is nearly complete, although the dissipation rate in the 
fast prescription
slows down once the plasma temperature becomes nonrelativistic.
In general, transients associated with the initial
conditions vanish rapidly. However, these transients
reveal that for some initial conditions the slow dissipation prescription 
is unphysical in that it fails to create entropy.  

Using the asymptotic solutions, the radius at which the
oscillating part of the magnetic field is completely dissipated 
can be estimated in terms of the magnetization parameter $\michelmu$ 
introduced by
\cite{michel69} and the dimensionless power of the flow
$\lhat$.
In the case of the Crab, it was shown by LK that slow reconnection 
cannot dissipate the magnetic energy 
inside the radius at which
observations imply that the wind terminates, independent
of the value of $\michelmu$. 
Faster dissipation prescriptions do not change this
conclusion, provided standard estimates of the 
rate at which the pulsar injects pairs into the Nebula 
are adopted \citep{hibschmanarons01a}.

However, our calculations show that a higher injection rate
permits faster dissipation. Thus, the $\sigma$-problem 
is resolved if dissipation is fast and $\mu \approx 2\times 10^4$, 
corresponding to a pair flux of 
$\Ndot = 3\times 10^{40} {\rm s}^{-1}$. This value is consistent with the
model 
of the synchrotron
spectrum of the Crab nebula presented by \citet{gallantetal02}.
\citet{lyubarskyeichler01} quote an even lower value of 
$\michelmu\approx10^3$
from their considerations of the jet emerging from the Crab pulsar. 
In this case, 
the $\sigma$-problem is also resolved by dissipation proceeding at the
tearing-mode limited rate. This suggests that most of the energy
emitted by the Crab pulsar as Poynting flux is converted into
particle-born energy during the supersonic expansion phase of the
wind. If a fraction of this power is converted into radiation, 
the dynamics of the flow imply that 
it will appear pulsed to an observer, which lends support to 
a recent suggestion concerning the origin of the optical to gamma-ray
pulses from this object \citep{kirkskjaeraasengallant02}.

\acknowledgments
We thank Yves Gallant and Yuri Lyubarsky for stimulating discussions
and helpful comments on this work.

\appendix
\section{The relativistic Harris equilibrium}
\label{harris}
The relativistic Vlasov equation:
\eqb
{\partial f\over \partial t} + \vec{v}\cdot{\partial f\over
 \partial \vec{r}}+\dot{\vec{p}}\cdot
{\partial f\over \partial \vec{p}}&=&0,
\label{relvlasov}
\eqe
where $\vec{v}$ is the particle 3-velocity, $\vec{p}=\gamma m\vec{v}$ the
momentum (with $\gamma=1/\sqrt{1-v^2/c^2}$), $f$ the Lorentz invariant
phase-space density,
\eqb
\dot{\vec{p}}&=&q\left(\vec{E} + {1\over c}\vec{v}\wedge\vec{B}\right)
\label{lorentzforce}
\eqe
the Lorentz force, $q$ the particle charge and $\vec{E}$ and 
$\vec{B}$ the electric and magnetic fields,
is solved by an arbitrary function of the constants of
motion.

In the case of a current sheet in the $y$-$z$ plane, where quantities in
(\ref{relvlasov}) depend on space only via $x$, the constants of motion are
the energy and the $y$ and $z$ components of the canonical momentum:
\eqb
E&=&\left(m^2 c^4 + c^2 p^2\right)^{1/2}
,\nonumber\\
P_y&=& p_y+{q\over c}A_y(x)
,\nonumber\\
P_z&=& p_z+{q\over c}A_z(x)
,\label{constants}
\eqe
where $\vec{A}$ is the vector potential. We 
seek a solution in which the $z$ component of $\vec{B}$
reverses at $x=0$. Choosing $A_z=0$, this means
$B_z=\partial A_y/\partial x=0$ at $x=0$.
To this we can add a constant component along $y$ by specifying 
$A_x=z B_y$, without affecting the following analysis. 

\citet{harris62} 
found a solution to this problem with a phase-space density which, at $x=0$, 
consists of two Maxwellians, one for each species, drifting in opposite
directions along the $y$ axis with zero charge density. 
The relativistic generalisation 
found by \cite{hoh68}
has two drifting  J\"uttner/Synge distributions. In the case
with just positrons $q=e$ and electrons $q=-e$, these distributions 
drift with equal and opposite speeds $c\beta q/e$
and have equal and opposite charge densities. 
In terms of quantities measured in the rest frame of a species
(denoted with a bar), 
the phase-space density 
of that component at $x=0$ is
\eqb
\bar{f}_{0\pm}&=&f_{0\pm}
\,=\,{\bar{N}\over 4\pi m^2c\Theta K_2\left(mc^2/\Theta\right)}{\rm
  exp}\left(-\bar{E}/\Theta\right),
\eqe
where $\bar{E}$ is the particle energy in the rest frame, 
$\Theta$ the temperature and $K_2$ a modified Bessel function.
The distribution is normalized such that the particle number density
is $\bar{N}$: 
$\int d^3p f = \bar{N}$.

Employing the Lorentz transformation 
\eqb
\bar{E}&=&\Gamma\left(E \mp c\beta p_y\right) ,
\eqe
where the upper sign refers to the positrons, 
we can write the phase-space density at $x=0$ 
in terms of lab.\ frame quantities:
\eqb
f_{0\pm}&=&{\bar{N}\over 4\pi m^2c\Theta K_2\left(mc^2/\Theta\right)}
{\rm exp}\left[ -{\Gamma\left( E\mp c\beta p_y\right)\over \Theta}\right]
.\label{relmax}
\eqe
Choosing $A_y(0)=0$, the solution of the Vlasov equation which matches
(\ref{relmax}) at $x=0$ is 
\eqb
f_{\pm}&=&{\bar{N}\over 4\pi m^2c\Theta K_2\left(mc^2/\Theta\right)}\times
\nonumber\\
&&{\rm exp}\left[ -{\Gamma\left( E\mp c\beta p_y\mp q\beta A_y(x)\right)\over 
\Theta}\right]
,\nonumber\\
&=&f_{0\pm}{\rm exp}\left[{e\beta\Gamma A_y(x)\over\Theta}\right]
.\label{relmax2}
\eqe
The current density carried by the electrons and positrons is:
\eqb
j_y&=&{\rm exp}\left[{e\beta\Gamma A_y(x)\over\Theta}\right]
ec^2\int d^3p {p_y \over E}\left(f_{0+} - f_{0-}\right).
\eqe
Using the invariance of $d^3p/E$:
\eqb
\int d^3p {p_y c^2\over E}f_{0\pm}&=&
\int d^3\bar{p} {p_y c^2\over \bar{E}}f_{0\pm}
,\nonumber\\
&=&\int d^3\bar{p} {c^2\over \bar{E}}\Gamma\left(\bar{p}_y\pm\beta \bar{E}/c\right)f_{0\pm}
,\nonumber\\
&=&
\pm\int d^3\bar{p} \Gamma\beta c f_{0\pm}\,=\,
\pm\Gamma\beta c \bar{N},
\eqe
so that
\eqb
j_y&=&2e\Gamma\beta c\bar{N}{\rm exp}\left[{e\beta\Gamma A_y(x)\over\Theta}\right].
\eqe
Using the Maxwell equation $4\pi j_y/c=-\partial B_z/\partial x$ we find:
\eqb
{\partial^2 A_y\over\partial x^2}&=&-8\pi\beta e N'_\pm{\rm exp}
\left[{e\beta\Gamma A_y\over \Theta}\right]
,\label{harrisclone}
\eqe
where $N'_\pm=\Gamma \bar{N}$ is the number density of each species in the lab.\ frame.

With the boundary conditions $\partial A_y/\partial x=A_y=0$ at $x=0$
the solution of (\ref{harrisclone}) is
\eqb
A_y&=&{\Theta\over e\beta\Gamma}\ln\left[\sech^2\left(x/a\right)\right]
,\nonumber\\
\noalign{\hbox{with}}\\
a&=&\ldebye/\beta
,\nonumber\\
\ldebye&=&\sqrt{\Theta\over 4\pi N'_\pm e^2\Gamma}.
\eqe
This is precisely the solution found by \citet{harris62}, with
the substitution $T=\Theta/\Gamma$. 
The distribution function is 
\eqb
f_\pm&=&f_{0\pm}\sech^2\left(x/a\right)
\eqe
and the density and magnetic field are
\eqb
n'_\pm&=&N'_\pm\sech^2\left(x/a\right)
,\nonumber\\
B'&=&\sqrt{16\pi N'_\pm\Theta\over\Gamma}\tanh\left(x/a\right).
\eqe
The number of particles per unit area of the sheet is
\eqb
N'_{\rm tot}&=&2\int_{-\infty}^\infty dx\, n'_\pm(x)\,=\,4 aN'_\pm.
\eqe

To calculate the elements of the stress-energy tensor
\eqb
T^{a,b}&=&\int {d^3p\over E} p^a p^b \left(f_+ + f_-\right),
\eqe
where $p^a=(E,cp)$ is the momentum 4-vector, 
it is convenient to transform to the rest frame of each species in turn 
and use the integrals
\eqb
\int d^3\bar{p} \bar{E} f_{0\pm}&=& \bar{N}\left[3\Theta+ mc^2 {K_1(mc^2/\Theta)
\over K_2(mc^2/\Theta)}\right]
,\nonumber\\
&\equiv&\bar{N}
R(mc^2/\Theta)
,\\
\int d^3\bar{p}{c^2 {\bar{p}}^2\over \bar{E}} f_{0\pm}&=& 3\bar{N}\Theta,
\eqe
to find:
\eqb
T^{0,0}&=&2n'_\pm\left[\Gamma\left(R+\Theta\right)-{\Theta\over \Gamma}\right]
\label{stress00},\\
T^{1,1}&=&T^{3,3}\,=\,2n'_\pm{\Theta\over\Gamma}
\label{stress11},\\
T^{2,2}&=& 2n'_\pm\left[\Gamma\left(R+\Theta\right)-{R\over\Gamma}\right]
\eqe
and all remaining elements zero.
In the pulsar wind, purely radial motion is considered, with the 
sheet normal along the radius vector. In this case, only the $T^{0,0}$ and
$T^{1,1}$ components are relevant and we can identify the effective enthalpy
density
and pressure: $w=2n'_\pm\Gamma\left(R+\Theta\right)$, $p=2n'_\pm\Theta/\Gamma$.
This leads to an effective ratio of specific heats 
[see \citet{lyubarskykirk01} Eq.~(A9)]:
\eqb
\gheat&=&1+{\Theta\over\Gamma\left(\Gamma R-mc^2\right)+\Gamma^2\beta^2\Theta}
\label{effspecificheat}.
\eqe
\section{The entropy condition
  Eq.~(\protect\ref{evolutioncondition})}
\label{appendixentropy}
Details of the derivation of eqs.~(\ref{continuity}), (\ref{energy})
and (\ref{entropy}) are given in LK. 
Using the same notation, we derive here the condition that
the specific entropy of a fluid element always increase,
Eq.~(\ref{evolutioncondition}). 

The fast phase variable $\phi$ and slow
radius variable $\rslow$, Eq.~(A15) of LK
are defined by
\eqb
\phi&=&\Omega\left[ t-\int_{r_0}^{r}{\diff r'\over
    v(r')
}\right]
,\nonumber\\
\rslow&=&\eps r/\rlight
\label{fastslowvariables}
\enspace,
\eqe
where $r_0$ is an arbitrary point in the flow.
The starting point of the derivation is the entropy
equation written in terms of the 
energy density $e'$ and entropy density $w'$ in the co-moving frame, 
this reads:
\eqb
\gamma\left(1-{v\over v_{\rm w}}\right){\partial e'\over\partial \phi}+
w'\left[{\partial\gamma\over\partial
    \phi}-{1\over\rslow^2v_{\rm w}}{\partial\over\partial\phi}\left(\gamma v
    \rslow^2\right)\right]&&
\nonumber\\
+\eps\gamma v{\partial e'\over\partial\rslow}+{\eps
  w'\over\rslow^2}{\partial\over\partial\rslow}
\left(\gamma v \rslow^2\right)
\,=\,{1\over\Omega}\gamma\left(E-{v\over c}B\right)j&&
\enspace.
\eqe
Expanding the dependent variables as in Eq.~(A18) of LK
and collecting terms of zeroth order one finds, using conditions
(A19--A22) of LK, that this equation is 
trivially satisfied, because the ohmic dissipative term $j'.E'$ vanishes to
zeroth order. To first order, we find an expression equivalent
to (A27) of LK:
\eqb
\lefteqn{-{\gamma_0 v_1\over v_0}{\partial e_0'\over\partial\phi}-{\gamma_0 w_0'\over
  v_0}{\partial v_1\over\partial\phi} +
\gamma_0 v_0{\partial
  e_0'\over\partial\rslow}+{w_0'\over\rslow^2}{\partial\over\partial\rslow}
\left(\gamma_0 v_0\rslow^2\right)}&&
\nonumber\\
&=&
{\gamma_0 v_1\over v_0}{\partial p_0\over\partial\phi} -
{B_0\over4\pi\gamma_0\rslow}{\partial\over\partial\rslow}\left(
\rslow v_0 B_0\right)
\enspace,
\eqe
in which the right hand-side is proportional to the ohmic dissipation.
To derive the entropy equation Eq.~(\ref{entropy}), this equation is
integrated over $\phi$ and exact periodicity is imposed upon the 
expanded dependent variables. However, in order to obtain
Eq.~(\ref{evolutioncondition}), we instead impose the restriction that
the ohmic dissipation be positive:
\eqb
\lefteqn{-{\gamma_0 v_1\over v_0}{\partial e_0'\over\partial\phi}-
{\gamma_0 w_0'\over
  v_0}{\partial v_1\over\partial\phi}}&&
\nonumber\\ 
&&+\gamma_0 v_0{\partial
  e_0'\over\partial\rslow}+{w_0'\over\rslow^2}{\partial\over\partial\rslow}
\left(\gamma_0 v_0\rslow^2\right)\,>\,0
\enspace.
\eqe
In order to integrate this condition, we introduce the ratio of
specific heats $\gheat$ as in Eq.~(\ref{specheats}) and assume this to be
constant. Multiplying by the
(positive) factor ${p'}^{-\left(\gheat-1\right)/\gheat}$,
integrating and imposing exact periodicity leads to:
\eqb
\left<{\partial\over\partial\rslow}\left(\rslow^2\gamma_0 v_0
      {p'}_0^{1/\gheat}\right)\right>&>&0
\enspace.
\eqe
Noting that for the Harris current sheet
\eqb
\left<{p'}_0^{1/\gheat}\right>
&=&\left({{\binfty}^2\over8\pi}\right)^{1/\gheat}\Delta
\left\lbrace
{1\over2}\int_{-\infty}^\infty\diff x\left[\sech(x)\right]^{2/\gheat}
\right\rbrace
\enspace,
\eqe
we arrive, after introducing the dimensionless variables used in
Sect.~\ref{sheet},
at condition (\ref{evolutioncondition}).
\section{Expansion speed of the current sheet}
\label{appendixspeed}
Assuming the sheet is bounded in the comoving frame by
a forward and a backward moving front at $r=r_+(r,t)$ and $r_-(r,t)$,
respectively, we can write their speeds
in the lab.\ frame as 
\eqb
{\diff r_\pm\over \diff t}&=&{v\pm c\vchar\over 1\pm v\vchar/c}
\enspace,
\eqe
where $\pm\vchar$ are the speeds of the fronts measured in an inertial 
frame moving with the speed of the wind at $(r,t)$. 
In terms of 
the fast phase and slow radius variables, Eq.~(\ref{fastslowvariables}),
we find
\eqb
{\diff \phi_\pm\over \diff\rslow}&=&
{ -\left(\diff r_\pm/\diff t\right)
\left(\partial t/\partial\rslow\right)+
\left(\partial r/\partial\rslow\right)
\over
-\left(\diff r_\pm/\diff t\right)
\left(\partial t/\partial\phi\right)+
\left(\partial r/\partial\phi\right)
}
\enspace.
\eqe
The fraction of
one wavelength occupied by the two sheets thus evolves according to 
\eqb
{\diff \Delta\over\diff
  \rslow}&=&{1\over\pi}{\diff\over\diff\rslow}
\left(\phi_+-\phi_-\right)
,\nonumber\\
&=&{2\vchar c^2\over\pi\eps\gamma^2\left(v^2-c^2\vchar^2\right)}
\enspace.
\eqe
This is equivalent to Eq.~(\ref{fixedspeed}), 
provided $r$ is identified as a \lq slow\rq\ variable.


\begin{thebibliography}{}
%
\bibitem[Arons(1983)]{arons83}
Arons J., 1983 in {\em Electron-Positron Pairs in Astrophysics}
 Eds.\ M.L.~Burns, A.K.~Harding, R.~Ramaty (New York: American Institute of
 Physics), p.~163
%
\bibitem[Begelman \& Li(1992)]{begelmanli92}
Begelman, M.C., Li Z.-Y., 1992 ApJ 397, 187
%
\bibitem[Beskin et al(1998)]{beskinkuznetsovarafikov98}
Beskin V.S., Kuznetsova I.V., Rafikov R.R., 1998 MNRAS 299, 341
%
\bibitem[Biskamp(2000)]{biskamp00}
Biskamp D., 2000 \lq\lq Magnetic reconnection in plasmas\rq\rq\ 
(Cambridge University Press, Cambridge)
%
\bibitem[Bogovalov(1999)]{bogovalov99}
Bogovalov S.V., 1999 A\&A 349, 1017
%
\bibitem[Bogovalov(2001)]{bogovalov01}
Bogovalov S.V., 2001 A\&A 371, 1155
%
\bibitem[Bogovalov \& Tsinganos(1999)]{bogovalovtsinganos99}
Bogovalov S.V., Tsinganos K. 1999 MNRAS 305, 211
%
\bibitem[B\"uchner \& Kuska(1999)]{buechnerkuska99}
B\"uchner J., Kuska J.P. 1999 Ann.\ Geophys.\ 17, 604
%
\bibitem[Buckley(1977)]{buckley77}
Buckley R. 1977 \mnras\ 180, 125
%
\bibitem[Chiueh, Li \& Begelman(1998)]{chiuehlibegelman98}
Chiueh T., Li Z.Y., Begelman M.C. 1998 \apj\ 505, 835
%
\bibitem[Coroniti(1990)]{coroniti90}
Coroniti F.V. 1990 ApJ 349, 538
%
\bibitem[Daughton(1999)]{daughton99}
Daughton W. 1999 Phys.\ Plasmas 6, 1329
%
\bibitem[Drenkhahn(2002)]{drenkhahn02}
Drenkhahn G. 2002 A\&A 387, 714
%
\bibitem[Drenkhahn \& Spruit(2002)]{drenkhahnspruit02}
Drenkhahn G., Spruit H.C. 2002 A\&A 391, 1141
%
\bibitem[Emmering \& Chevalier(1987)]{emmeringchevalier87}
Emmering R.T., Chevalier R.A. 1987
ApJ 321, 334
%
\bibitem[Gallant et al(2002)]{gallantetal02}
Gallant Y.A., van der Swaluw E., Kirk J.G., Achterberg A. 2002
in \lq\lq Neutron Stars in Supernova Remnants\rq\rq, 
Eds: P.O. Slane, B.M. Gaensler,
ASP Conference Series Vol.\ 271, 99
%
\bibitem[Gunn \& Ostriker(1969)]{gunnostriker69}
Gunn J.E., Ostriker J.P., 1969 Nature 221, 454
%
\bibitem[Harris(1962)]{harris62}
Harris E.G. 1962 Nuovo Cimento 23, 115
%
\bibitem[Heinz \& Begelman(2000)]{heinzbegelman00}
Heinz S., Begelman M.C. 2000 ApJ 535, 104
%
\bibitem[Hesse et al.(2001)]{hesseetal01}
Hesse M., Kuznetsova M., Birn J. 2001
J.\ Geophys. Res. Space Phys. 106, 29831
%
\bibitem[Hibschman \& Arons(2001a)]{hibschmanarons01a}
Hibschman J.A., Arons J. 2001a ApJ 554, 624
%
\bibitem[Hibschman \& Arons(2001b)]{hibschmanarons01b}
Hibschman J.A., Arons J. 2001b ApJ 560, 871
%
\bibitem[Hoh(1968)]{hoh68}
Hoh F.C. 1968 Physics Fluids 9, 277
%
\bibitem[Contopoulos \& Kazanas(2002)]{contopouloskazanas02}
Contopoulos I., Kazanas D. 2002 \apj\ 566, 336
%
\bibitem[Kennel \& Coroniti(1984)]{kennelcoroniti84}
Kennel C.F., Coroniti F.V., 1984 \apj\ 283, 694
%
\bibitem[Kirk \& Duffy(1999)]{kirkduffy99}
Kirk J.G., Duffy P. 1999 J.\ Phys.\ G: Nucl.\ Part.\ Phys.\ 25, R163
%
\bibitem[Kirk et al(2002)]{kirkskjaeraasengallant02}
Kirk J.G., Skj\ae raasen O., Gallant Y.A. 2002 A\&A 388, L29
%
\bibitem[Kuijpers(2001)]{kuijpers01}
Kuijpers J. 2001 Publ\ Astron.\ Soc.\ Aust.\ 18, 407
%
\bibitem[Lovelace et al(2002)]{lovelaceetal02}
Lovelace R.V.E., Li H., Koldoba A.V., Ustyugova G.V., Romanova, M.M. 2002
ApJ 572, 445
%
\bibitem[Lyubarsky(1996)]{lyubarsky96}
Lyubarsky Y. 1996 A\&A 311, 172 
%
\bibitem[Lyubarsky(2002a)]{lyubarsky02a}
Lyubarsky Y. 2002a MN 329, L34 
%
\bibitem[Lyubarsky(2002b)]{lyubarsky02b}
Lyubarsky Y. 2002b MN in press, astro-ph/0211046 
%
\bibitem[Lyubarsky \& Eichler(2001)]{lyubarskyeichler01}
Lyubarsky Y., Eichler D. 2001 ApJ 562, 494 
%
\bibitem[Lyubarsky \& Kirk(2001)]{lyubarskykirk01}
Lyubarsky Y., Kirk J.G. 2001 ApJ 547, 437 (LK in text) 
%
\bibitem[Lyutikov \& Blackman(2001)]{lyutikovblackman01}
Lyutikov M., Blackman E.G. 2001 MN, 321, 177
%
\bibitem[Lyutikov \& Uzdensky(2002)]{lyutikovuzdensky02}
Lyutikov M., Uzdensky D. 2002 astro-ph/0210206
%
\bibitem[Melatos(2002)]{melatos02}
Melatos A. 2002 in \lq Neutron Stars and Supernova Remnants\rq\ ASP Conf.\
Series Vol.\ 271, Eds: P.O. Slane and B.M. Gaensler, ASP, San Francisco, page 115
%
\bibitem[Melatos \& Melrose(1996)]{melatosmelrose96}
Melatos A., Melrose D.B. 1996 MN 279, 1168
%
\bibitem[Melrose(1986)]{melrose86}
Melrose D.B. 1986 \lq\lq Instabilities in space and laboratory  
plasmas\rq\rq, Cambridge University Press, Cambridge
%
\bibitem[Michel(1969)]{michel69}
Michel, F.C., 1969 ApJ 158, 727
%
\bibitem[Michel(1982)]{michel82}
Michel, F.C., 1982 Rev.\ Mod.\ Phys.\ 54, 1
%
\bibitem[Michel(1994)]{michel94}
Michel, F.C., 1994 ApJ 431, 397
%
\bibitem[Priest \& Forbes(2000)]{priestforbes00}
Priest E.R., Forbes T. 2000 \lq\lq Magnetic Reconnection\rq\rq\ (Cambridge
University Press, Cambridge)
%
\bibitem[Pritchett et al.(1996)]{pritchettetal96}
Pritchett P.L., Coroniti F.V., Decyk V.K. 1996 
J.\ Geophys. Res. Space Phys. 101, 27413
%
\bibitem[Rees \& Gunn(1974)]{reesgunn74}
Rees M.J., Gunn J.E., 1974 \mnras\ 167, 1
%
\bibitem[Schopper et al(1998)]{schopperetal98}
Schopper R., Lesch H., Birk G.T. 1998
A\&A 335, 26
%
\bibitem[Skj\ae raasen \& Kirk(2001)]{skjaeraasenkirk01}
Skj\ae raasen O., Kirk J.G. 2001 in \lq\lq Similarities and 
universality in relativistic flows\rq\rq\ Eds: Georganopoulos, Guthmann, 
Manolakou, Marcowith (Logos Verlag, Berlin) 
%
\bibitem[Skj\ae raasen et al.(2002)]{skjaeraasenetalboston02}
Skj\ae raasen O., Kirk J.G., Y.A. Gallant 2002
in \lq Neutron Stars and Supernova Remnants\rq\ ASP Conf.\
Series Vol.\ 271, Eds: P.O. Slane and B.M. Gaensler, ASP, San Francisco, page 87
%
\bibitem[Somov \& Titov(1985)]{somovtitov85}
Somov B.V., Titov V.S. 1985 Solar Physics 102, 79
%
\bibitem[Spruit(1999)]{spruit99}
Spruit H.C. 1999
A\&A 341, L1
%
\bibitem[Spruit et al(2001)]{spruitdaignedrenkhahn01}
Spruit H.C., Daigne, F., Drenkhahn G. 2001
A\&A 369, 694
%
\bibitem[Thompson(1994)]{thompson94}
Thompson C. 1994 MN 270, 480
%
\bibitem[Usov(1975)]{usov75}
Usov V.V., 1975 Astrophys.\ \& Space Sci.\ 32, 375
%
\bibitem[Usov(1992)]{usov92}
Usov V.V., 1992 Nature 357, 472
%
\bibitem[Weisskopf et al(2000)]{weisskopfetal00}
Weisskopf M., Hester J.J., Tennant A.F., Elsner R.F., Schulz N.S., Marshall
H.L., Karovska M., Nichols J.S., Swartz D.A., Kolodziejczak J.J., O'Dell S.L. 2000
ApJ 536, L81
%
\bibitem[Zelenyi \& Krasnosel'skikh(1979)]{zelenyikrasnoselskikh79}
Zelenyi L.M., Krasnosel'skikh V.V. 1979 Sov. Phys. A.J. 56, 819
%
\bibitem[Zenitani \& Hoshino(2001)]{zenitanihoshino01}
Zenitani S., Hoshino M. 2001 ApJ 562, L63
%
\bibitem[Zhu \& Winglee(1996)]{zhuwinglee96}
Zhu Z.W., Winglee R.M. 1996 J.\ Geophys. Res. Space Phys. 101, 4885

\end{thebibliography}
\end{document}